\documentclass{aa}  
\usepackage{amsmath}
\usepackage[english]{babel}           
\usepackage{graphicx,epsfig}
\usepackage{txfonts}
\usepackage{lscape}
\usepackage{eucal}

\newcommand{\etal}{{\it{et al.~}}}

\def\cc{\ifmmode{\,{\rm cm}^{-3}}\else{$\,{\rm cm}^{-3}$}\fi}
\def\cq{\ifmmode{\,{\rm cm}^{-2}}\else{$\,{\rm cm}^{-2}$}\fi}
\def\mic{\ifmmode{\,\mu{\rm m}}\else{$\mu$m}\fi}
\def\eccs{\ifmmode{\,{\rm erg}\,{\rm cm}^{-3} {\rm s}^{-1}}\else{$\,{\rm
erg}\,{\rm cm}^{-3} {\rm s}^{-1}$}\fi}
\def\ecqs{\ifmmode{\,{\rm erg}\,{\rm cm}^{-2}\,{\rm s}^{-1}\,{\rm
sr}^{-1}}\else{$\,{\rm erg}\,{\rm cm}^{-2}\,{\rm s}^{-1}\,{\rm sr}^{-1}$}\fi}
\def\deg{\ifmmode{^{\circ}}\else{$^{\circ}$}\fi} 
\def\pc{\ifmmode{\,{\rm pc}}\else{$\,{\rm pc}$}\fi} 
\def\kms{\ifmmode{\,{\rm km}\,{\rm s}^{-1}}\else{km s$^{-1}$}\fi} 
\def\kmspc{\ifmmode{\,{\rm km}\,{\rm s}^{-1}\,{\rm pc}^{-1}}\else{km
s$^{-1}$ pc$^{-1}$}\fi} 
\def\MJysr{\ifmmode{\,{\rm MJy\,sr}^{-1}}\else{$\,{\rm MJy\,sr}^{-1}$}\fi} 
\def\Kkms{\ifmmode{\,{\rm K\,km\,s}^{-1}}\else{$\,{\rm K\,km\,s}^{-1}$}\fi} 
\def\twCO{\ifmmode{\rm ^{12}CO}\else{$\rm^{12}CO$}\fi} 
\def\thCO{\ifmmode{\rm ^{13}CO}\else{$\rm^{13}CO$}\fi} 
\def \Cp{\ifmmode{\rm C^+}\else{$\rm C^+$}\fi} 
\def \CHp{\ifmmode{\rm CH^+}\else{$\rm CH^+$}\fi}
\def \CHtp{\ifmmode{\rm CH_2^+}\else{$\rm CH_2^+$}\fi} 
\def \CtH{\ifmmode{\rm C_2H}\else{$\rm C_2H$}\fi} 
\def \CthHt{\ifmmode{\rm C_3H_2}\else{$\rm C_3H_2$}\fi} 
\def\CHthp{\ifmmode{\rm CH_3^+}\else{$\rm CH_3^+$}\fi} 
\def \HCOp{\ifmmode{\rm HCO^+}\else{$\rm HCO^+$}\fi} 
\def \HtOp{\ifmmode{\rm H_3O^+}\else{$\rm H_3O^+$}\fi} 
\def \HCfiN{\ifmmode{\rm HC_5N}\else{$\rm HC_5N$}\fi} 
\def\wat{\ifmmode{\rm H_2O}\else{$\rm H_2O$}\fi} 
\def \oxy{\ifmmode{\rm O_2}\else{$\rm O_2$}\fi} 
\def \HH{\ifmmode{\rm H_2}\else{$\rm H_2$}\fi}
\def \Jone{\ifmmode{\rm {(J=1--0)}}\else{{(J=1--0)}}\fi} 
\def\Jtwo{\ifmmode{\rm {(J=2--1)}}\else{{(J=2--1)}}\fi} 
\def\Jthr{\ifmmode{\rm {(J=3--2)}}\else{{(J=3--2)}}\fi} 
\def\Jfou{\ifmmode{\rm {(J=4--3)}}\else{{(J=4--3)}}\fi} 
\def\Jfiv{\ifmmode{\rm {J=4--3}}\else{{J=4--3}}\fi} 
\def \Ta{\ifmmode{\rm T_A}\else{$\rm T_A$}\fi} 
\def \Tas{\ifmmode{\rm T_A^*}\else{$\rm T_A^*$}\fi} 
\def \Tmb{\ifmmode{\rm T_{mb}}\else{$\rm T_{mb}$}\fi} 
\def \Tr{\ifmmode{\rm T_r}\else{$\rm T_r$}\fi} 
\def \Trs{\ifmmode{\rm T_r^*}\else{$\rm T_r^*$}\fi}

\begin{document}

\title{Molecular absorption lines toward star-forming regions : a comparative study
    of HCO$^{+}$, HNC, HCN, and CN\thanks{Based on observations obtained with
    the IRAM 30m telescope. IRAM is supported by INSU/CNRS (France), MPG
    (Germany), and IGN (Spain).}}

\author{\vspace{2cm}
  B. Godard \inst{1,2}, 
  E. Falgarone \inst{1},
  M. Gerin \inst{1},
  P. Hily-Blant \inst{3},
  \and 
  M. De Luca \inst{1}
}

\institute{LRA/LERMA, CNRS UMR 8112, Observatoire de Paris \& \'Ecole Normale
  Sup\'erieure, Paris
  \and 
  IAS, CNRS UMR 8617, Universit\'e Paris-Sud, Orsay
  \and
  LAOG, CNRS UMR 5571, Universit\'e Joseph Fourier \& Observatoire de Grenoble, Grenoble}

 \date{Received 18 February 2010 / Accepted 25 May 2010}

\abstract{}
{The comparative study of several molecular species at the origin of the
  gas phase chemistry in the diffuse interstellar medium (ISM) is a key input
  in unraveling the coupled chemical and
  dynamical evolution of the ISM.}
{
The lowest rotational lines of HCO$^{+}$, HCN, HNC, and CN were observed
at the IRAM-30m telescope in absorption against the $\lambda 3$ mm and
$\lambda 1.3$ mm continuum emission of massive star-forming regions in the
Galactic plane. The absorption lines probe  the  gas over kiloparsecs along
these lines of sight. The excitation temperatures of HCO$^{+}$ are
inferred from the comparison of the  absorptions in the two lowest
transitions. The spectra of all molecular species on the same line of sight are
decomposed  into Gaussian velocity components. Most appear in all the spectra
of a given line of sight. For each component, we derived the central
opacity, the velocity dispersion, and computed the molecular
  column density. We compared our results to the predictions of 
  UV-dominated chemical models of photodissociation regions (PDR models) and to
  those of non-equilibrium models in which the chemistry is driven
  by the dissipation of turbulent energy (TDR models).}
{The molecular column densities of all the velocity components span up to two
  orders of magnitude. Those of CN, HCN, and HNC are linearly correlated with
  each other with mean ratios 
$N({\rm HCN})/ N({\rm HNC}) = 4.8 \pm 1.3$ and
$N({\rm CN })/ N({\rm HNC}) = 34 \pm 12$,
and more loosely correlated with those of HCO$^{+}$, 
$N({\rm HNC}) / N({\rm HCO}^{+}) = 0.5  \pm 0.3$ ,
$N({\rm HCN}) / N({\rm HCO}^{+}) = 1.9 \pm 0.9$, and
$N({\rm CN }) / N({\rm HCO}^{+}) = 18 \pm 9$.
  These ratios are similar to those inferred from observations of high 
  Galactic latitude lines of sight, suggesting that the gas sampled by
  absorption lines  in the Galactic plane has
  the same chemical properties as that in the Solar neighbourhood. 
The FWHM of the Gaussian velocity components span the range 0.3 to 3 \kms\ 
and those of the HCO$^{+}$ lines are found to be 30\% broader than those of
CN-bearing molecules.
  The PDR models fail to reproduce simultaneously the observed abundances of
  the CN-bearing species and HCO$^{+}$, even for high-density material ($
  100$ cm$^{-3}$ $< n_{\rm H} < 10^{4}$ cm$^{-3}$). The TDR models, in turn,
  are able to reproduce the observed abundances and abundance ratios of all
  the analysed molecules for the moderate gas densities ($ 30$ cm$^{-3}$ $<
  n_{\rm H} < 200$ cm$^{-3}$) and the turbulent energy observed in the diffuse
  interstellar medium.} 
{Intermittent turbulent dissipation appears to be a promising driver of the
  gas phase chemistry
of the diffuse and translucent gas throughout the Galaxy. The details of the
dissipation  mechanisms still need to be investigated.}

   \keywords{Astrochemistry - Turbulence - ISM: molecules -
     ISM: kinematics and dynamics - ISM: structure - ISM: clouds
               }

   \authorrunning{B. Godard \etal}
   \titlerunning{Molecular absorption lines toward star-forming regions} 
   \maketitle
%

\section{Introduction}

\begin{table*}[!!!ht]
\begin{center}
\caption{Properties of background sources, and rms noise levels of the spectra.}
\begin{tabular}{l @{\hspace{0.3cm}} l @{\hspace{0.3cm}} l @{\hspace{0.3cm}} l
    @{\hspace{0.3cm}} l @{\hspace{0.3cm}} l @{\hspace{0.3cm}} c
    @{\hspace{0.3cm}} c @{\hspace{0.3cm}} c @{\hspace{0.3cm}} c
    @{\hspace{0.3cm}} c}
\hline
Source      & RA(J2000)   & Dec (J2000) & l & b & D$^{a}$ & 
$\sigma_l/T_{\rm c}$ $^{b}$ & $\sigma_l/T_{\rm c}$ $^{b}$ & $\sigma_l/T_{\rm c}$
    $^{b}$ & $\sigma_l/T_{\rm c}$ $^{b}$ & $\sigma_l/T_{\rm c}$ $^{b}$ \\
            & (h) (m) (s) & ($^{\circ}$) ($'$) ($''$) &  ($^{\circ}$) & ($^{\circ}$) & (kpc) &
HCO$^{+}$ (0-1) & HCO$^{+}$ (1-2) & HNC (0-1) & HCN (0-1) & CN (0-1)  \\
\hline
G05.88-0.39 & 18 00 30.4 & -24 04 00   & 05.88 & -0.39 & 3.8  &
0.021       &            &             &               &       \\
G08.67-0.36 & 18 06 18.9 & -21 37 35   & 08.67 & -0.36 & 4.8  & 
0.138       &            &             &               &       \\
G10.62-0.38 & 18 10 28.7 & -19 55 50   & 10.62 & -0.38 & 4.8  & 
0.049       & 0.206      & 0.043       & 0.035         & 0.064 \\
G34.3+0.1   & 18 53 18.7 & +01 14 58   & 34.26 & +0.15 & 3.8  & 
0.117       & 0.328      & 0.038       & 0.036         & 0.154  \\
W49N        & 19 10 13.2 & +09 06 12   & 43.17 & +0.01 & 11.5 & 
0.015       & 0.072      & 0.020       & 0.021         & 0.031 \\
W51         & 19 23 43.9 & +14 30 30.5 & 49.49 & -0.39 & 7.0  & 
0.022       & 0.089      & 0.043       & 0.034         & 0.070 \\
\hline
\end{tabular}
\begin{list}{}{}
\item[$^{a}$] Source distance from  Fish \etal (2003) who resolve the kinematic
  distance ambiguity. Errors are around 1.0 kpc.
\item[$^{b}$] $\sigma_l/T_{\rm c}$ is the rms noise divided by the continuum
  intensity of the spectra, at a frequency resolution of 40 kHz.
\end{list}
\label{TabSources}
\end{center}
\end{table*}

\begin{table*}
\begin{center}
\caption{Observation parameters.}
\begin{tabular}{l @{\hspace{0.2cm}} c @{\hspace{0.2cm}} r @{\hspace{0.2cm}}
                c @{\hspace{0.2cm}} c @{\hspace{0.2cm}} c @{\hspace{0.2cm}}
                c @{\hspace{0.2cm}} c @{\hspace{0.2cm}} c @{\hspace{0.5cm}}
                c @{\hspace{0.2cm}} c @{\hspace{0.2cm}} c @{\hspace{0.2cm}}
                c @{\hspace{0.2cm}} c @{\hspace{0.2cm}} c}
\hline
Molecule & Transition  & \multicolumn{1}{c}{$\nu_0$ (GHz)} &
\multicolumn{6}{c}{$T_{\rm c}$ (K)} & \multicolumn{6}{c}{$T_{\rm sys}$ (K)} \\
 &  &  &  G05 & G08 & G10 & G34 & W49 & W51 & G05 & G08 & G10 & G34 & W49 & W51 \\
\hline
HCO$^{+}$ & 0 - 1                 &  89.1885247 & 1.5 & 0.22 & 0.69 & 0.86 & 1.73 & 1.03 & 156 & 162 & 206 & 137 & 167 & 132\\
HCO$^{+}$ & 1 - 2                 & 178.3750563 &     &      & 0.60 & 0.26 & 1.10 & 0.96 &     &     & 588 & 420 & 375 & 374\\
HNC       & 0,1 - 1,0             &  90.6634170 &     &      & 0.97 & 0.88 & 1.94 & 0.93 &     &     & 186 & 162 & 160 & 191\\
HNC       & 0,1 - 1,2             &  90.6635560 &     &      & 0.97 & 0.88 & 1.94 & 0.93 &     &     & 186 & 162 & 160 & 191\\
HNC       & 0,1 - 1,1             &  90.6636220 &     &      & 0.97 & 0.88 & 1.94 & 0.93 &     &     & 186 & 162 & 160 & 191\\
HCN       & 0,1 - 1,1             &  88.6304160 &     &      & 1.05 & 1.20 & 1.86 & 1.10 &     &     & 158 & 167 & 146 & 140\\
HCN       & 0,1 - 1,2             &  88.6318470 &     &      & 1.05 & 1.20 & 1.86 & 1.10 &     &     & 158 & 167 & 146 & 140\\
HCN       & 0,1 - 1,0             &  88.6339360 &     &      & 1.05 & 1.20 & 1.86 & 1.10 &     &     & 158 & 167 & 146 & 140\\
CN        & 0,1/2,3/2 - 1,1/2,1/2 & 113.1441573 &     &      & 0.94 & 0.70 & 1.68 & 0.92 &     &     & 280 & 382 & 275 & 275\\
CN        & 0,1/2,1/2 - 1,1/2,3/2 & 113.1704915 &     &      & 0.94 & 0.70 & 1.68 & 0.92 &     &     & 280 & 382 & 275 & 275\\
CN        & 0,1/2,3/2 - 1,1/2,3/2 & 113.1912787 &     &      & 0.94 & 0.70 & 1.68 & 0.92 &     &     & 280 & 382 & 275 & 275\\
CN        & 0,1/2,1/2 - 1,3/2,3/2 & 113.4881202 &     &      & 0.93 & 0.69 & 1.68 & 0.93 &     &     & 284 & 312 & 209 & 206\\
CN        & 0,1/2,3/2 - 1,3/2,5/2 & 113.4909702 &     &      & 0.93 & 0.69 & 1.68 & 0.93 &     &     & 284 & 312 & 209 & 206\\
CN        & 0,1/2,1/2 - 1,3/2,1/2 & 113.4996443 &     &      & 0.93 & 0.69 & 1.68 & 0.93 &     &     & 284 & 312 & 209 & 206\\
CN        & 0,1/2,3/2 - 1,3/2,3/2 & 113.5089074 &     &      & 0.93 & 0.69 & 1.68 & 0.93 &     &     & 284 & 312 & 209 & 206\\
CN        & 0,1/2,3/2 - 1,3/2,1/2 & 113.5204315 &     &      & 0.93 & 0.69 & 1.68 & 0.93 &     &     & 284 & 312 & 209 & 206\\
\hline
\end{tabular}
\label{TabCondObs}
\end{center}
\end{table*}

Since its discovery through absorption lines in bright star spectra
(Hartmann 1904), our knowledge of the interstellar medium has grown
thanks to a variety of absorption line measurements. Diatomic molecules
have been discovered in this hostile environment in the late 1930's and early 1940's
(see references in the review of Snow and McCall 2006) and since then
have been observed in the UV and visible spectral domains toward bright stars
(e.g. Crane \etal 1995, Gredel 1997, Weselak \etal 2008a, 2008b,
2009, Gry \etal 2002, Lacour \etal 2005) at increasingly high
extinction values (up to $A_V \leqslant 5$, Gredel \etal 2002).  Molecules are
also detected at submillimetre, millimetre, and centimetre wavelengths
in absorption against the continuum emission of star-forming regions (e.g. Koo
1997, Fish \etal 2003,
Nyman 1983, Nyman \& Millar 1989, Cox \etal 1988, Carral \& Welch
1992, Greaves \& Williams 1994, Neufeld \etal 2002, Plume et
al. 2004, Olofsson \etal 2010 in prep.) and bright extragalactic radio sources
(Liszt \etal 2008 and references therein).  The picture that emerges
from these measurements is complex, and the link between the structure
in density and temperature, the molecular richness, and the velocity
field still needs to be unraveled. In particular, the abundance of
several molecules, such as CH$^{+}$ or HCO$^{+}$, are found to be at least one
order of magnitude larger than those deduced 
from UV-dominated chemical models. In the cases of HCO$^{+}$, HNC, and
HCN the abundances inferred from observations of the diffuse ISM are
similar to those observed in dark clouds (Lucas \& Liszt 1996, 2000,
Liszt \& Lucas 2001). The column densities of specific species, like
HCO$^{+}$ and OH, exhibit remarkable correlations that cannot be understood in
the UV-dominated chemistry.

Alternative chemical models that couple the chemical evolution of the
gas to the turbulent dynamical evolution of the medium have been
developed. The space time intermittency of turbulent dissipation is
invoked to locally enhance the rate of highly endoenergetic reactions,
otherwise blocked in the cold ISM. Turbulent dissipation in low-velocity, 
magneto hydrodynamical (MHD) shocks (Flower \& Pineau des For\^ets
1998) or magnetized regions of intense velocity shear (Falgarone et
al. 1995, Joulain \etal 1998) are promising frameworks.  Turbulent
mixing between the warm neutral medium (WNM) and the cold neutral
medium (CNM) has also been proposed as a possibility to enhance 
the formation rate of specific species (Lesaffre \etal 2007).  In the
turbulent dissipation regions (TDR) model of Godard \etal (2009),
dissipation of turbulent energy occurs in short-lived ($\sim 10^{2}$
yr) magnetized vortices and is responsible for long-lasting ($\sim
10^{3}$ yr or more) chemical signatures due to the strong thermal and
chemical inertia of the diffuse gas. Following the chemical and
thermal evolutions of the dissipation and relaxation phases, 
the TDR models reproduce the column densities of
CH$^{+}$, CH, HCO$^{+}$, OH, H$_2$O, C$_2$H, and of the rotational
levels of H$_2$ ($J \geqslant 3$), as well as their correlations,
observed in the local diffuse medium.

The present study broadens the investigation of the local diffuse medium
chemistry by analysing observations performed in the direction of remote
star-forming regions that sample gas throughout the Galactic plane.  The 
selected background sources are extensively studied star-forming
regions, close to the Galactic plane ($|b| < 0.5^{\circ}$)
(e.g. Mookerjea \etal 2007).  These lines of sight are also the
primary targets of the Herschel-HIFI (Heterodyne Instrument for the
Far Infrared) key programme PRISMAS (PRobing InterStellar Molecules with
Absorption line Studies) whose objective is to advance the
understanding of astrochemistry by observing species at the origin of
gas phase chemistry, namely light hydrides and small molecules containing
carbon.  The present work is therefore anticipating future comparisons
between radio and far infrared (FIR) observations.

The observations and their analysis are discussed in Sects. \ref{SectObs} and
\ref{SectAna}. The main results are presented in Sect. \ref{SectRes} and 
the cyanide chemistry and  the predictions of 1D chemical
models are discussed  in Sect. \ref{SectDis}.



\section{Observations} \label{SectObs}

\begin{figure*}[!ht]
\begin{center}
\includegraphics[width=19.0cm,angle=0]{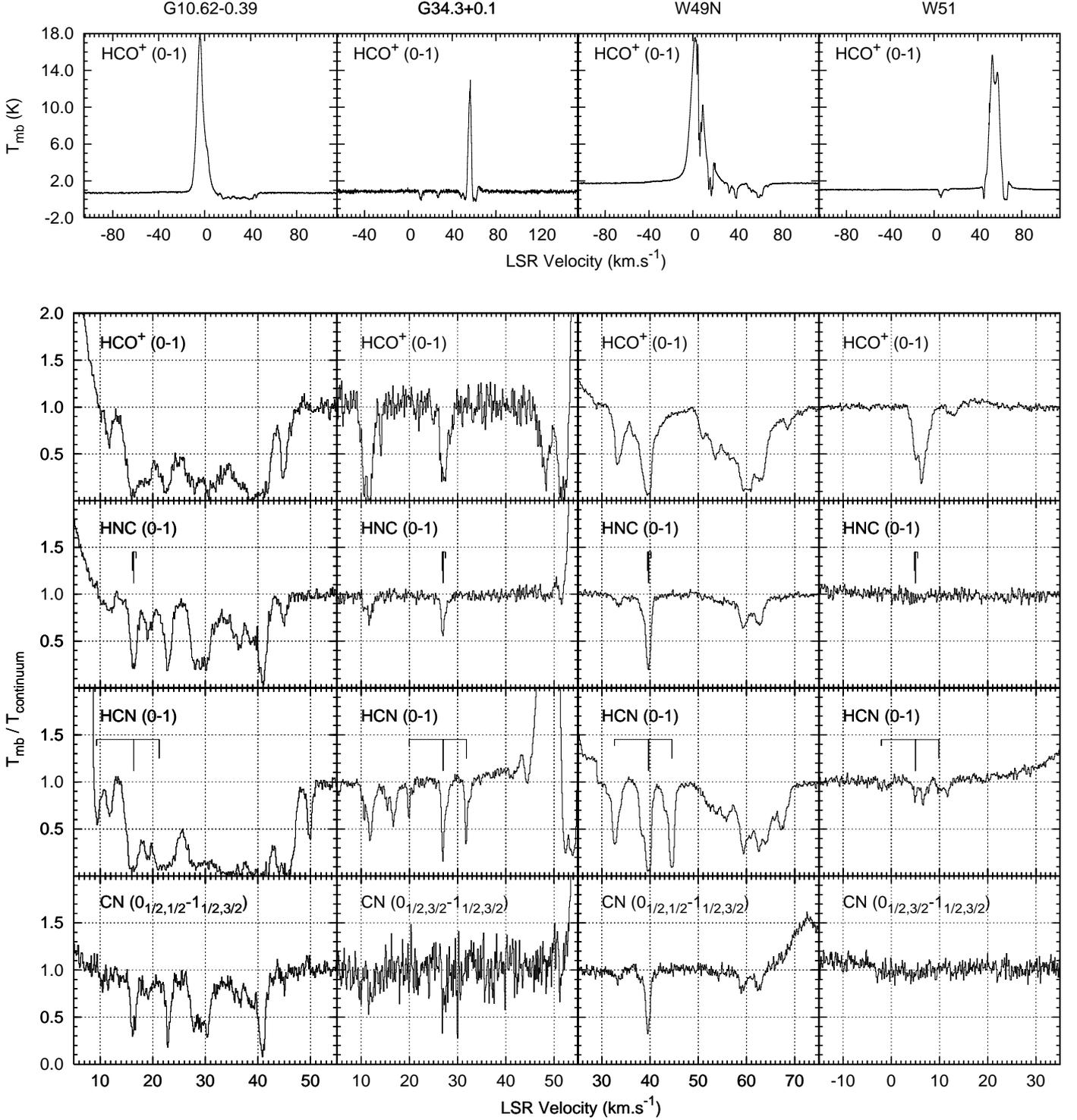}
\caption{Absorption profiles observed in the direction of  G10.62-0.39,
  G34.3+0.1,
  W49N, and W51 in the ground state transitions of HCO$^+$, HCN, HNC, and CN. 
  The hyperfine structure (relative positions and
  relative LTE line strength) of the $J=0-1$ transition of HNC and HCN are
  displayed. The broad velocity coverage of the upper panels illustrates the
  quality of the baseline and the complexity of the emission and
  absorption mixture. The lower panels display the
  velocity structure of the absorption features in more details.
  All the spectra of the lower panels have been normalized to the
  continuum temperature.}
\label{FigSpectres}
\end{center}
\end{figure*}

The observations were carried out at the IRAM-30m telescope at Pico Veleta
(Spain) in August and December 2006. 
For the sources listed in
Table \ref{TabSources} (with their Galactic coordinates and their distance from
the Sun), we observed in wobbler-switching mode:
\begin{itemize}
\item[$\bullet$] the $J=0-1$ absorption lines of HCO$^{+}$ and HNC,
\item[$\bullet$] the $F=1-1$, $1-2$ and $1-0$ hyperfine components of the
  $J=0-1$ absorption line of HCN, and
\item[$\bullet$] the $F=3/2-1/2$, $1/2-3/2$, $3/2-3/2$, $3/2-5/2$, and $1/2-1/2$
  hyperfine components of the $J=0-1$ absorption line of CN.
\end{itemize}
We used the IRAM-30m SIS receivers tuned in SSB (single side band) mode, with
rejection of the image bands larger than 15dB. The spectra were obtained
using the VESPA correlator at a frequency resolution of $40$ kHz.
The rms pointing accuracy of the telescope checked on nearby continuum sources
was 3$''$ and the integration time ranged between 11 and 59 minutes. Between
88 and 113 GHz, the forward and main beam efficiencies were of 0.95 and 0.75,
the spectral resolution and the half power beam width range from $\sim 0.13$
\kms\ and 28$''$ to $\sim 0.10$ \kms\ and 22$''$ respectively. The observation
parameters are listed in Table \ref{TabCondObs}.

The gain receiver stability, and therefore the uncertainty on the continuum
level, $T_{\rm c}$, 
is estimated by comparing the continuum levels from several spectra 
in the vicinity of the  different lines. For the same frequency ranges, they
differ by 0.1 to 0.3 K (see Table \ref{TabCondObs}).  The uncertainty on the
continuum level therefore ranges between 10 and 30 \%.

Fig. \ref{FigSpectres} displays a selection of 
spectra obtained after data reduction using the GILDAS-CLASS90
software\footnote{See http://www.iram.fr/IRAMFR/GILDAS for more information
  about GILDAS softwares.} (Hily-Blant \etal 2005). We focus
on the absorption part of the spectra, outside the line emission of the
star-forming region, since we are interested in the velocity structure and  
properties of the absorbing gas. In several cases, in order to extract as
much information on the absorption lines as possible, the emission in wings of
strong lines from the background source was removed using polynomial and
exponential fitting routines.



As shown in Fig \ref{FigSpectres}, the absorption spectra are highly
structured. While all the background 
sources are in the Galactic plane along lines of sight that cross the
molecular ring and the Sagittarius spiral arm once (except that toward W49N which
crosses it twice), the number of absorption features is highly variable from one
source to another: ranging from 1 to 20 components.
Optical depths, defined as
\begin{equation}
\tau = - {\rm ln} \left( \frac{T}{T_{\rm c}} \right) ,
\end{equation}
are also highly variable from one species to another; several components are
only detectable in HCO$^{+}$, and the CN spectra are unequivocally
less overcast than all the other spectra.
Complementary data have been obtained with the PdBI interferometer 
in the direction of W49N and W51 (Pety et
al., in preparation). 

The hyperfine components of the $J=0-1$ 
absorption lines of HNC (Bechtel \etal 2006) are too close to be individually
resolved given the significant velocity dispersion of the gas. 
The hyperfine structure causes a systematic broadening that depends on
the FWHM (full width at half maximum) $\Delta \upsilon_{\rm real}$ of the
velocity component: the broadening  $\Delta \upsilon - \Delta \upsilon_{\rm
  real}$ ranges\footnote{This
  result on the line profile broadening is derived from the analysis of 560
  synthetic spectra taking into account the hyperfine structure of HNC
  (line strength and velocity structure).} between 0.19 and 
0.035 \kms\ for line FWHM varying from 0.3 to 3 \kms.
In the case of HCN and CN (0-1), the separation of the hyperfine components is larger 
than most linewidths 
but the large number of velocity features along the lines of sight G10.62-0.38 and W49N induces
a blending of the hyperfine components.


While absorption features above the noise level appear at all widths,
down to the velocity resolution, we have not analysed the spectra to that
level: every feature we detect spans at least 3 spectrometer channels,
between 0.3 and 0.4 km s$^{-1}$ depending on the frequency.

\section{Analysis} \label{SectAna}

\subsection{Excitation temperatures} \label{SectTemp}
\begin{figure}[!ht]
\begin{center}
\includegraphics[width=9.0cm,angle=0]{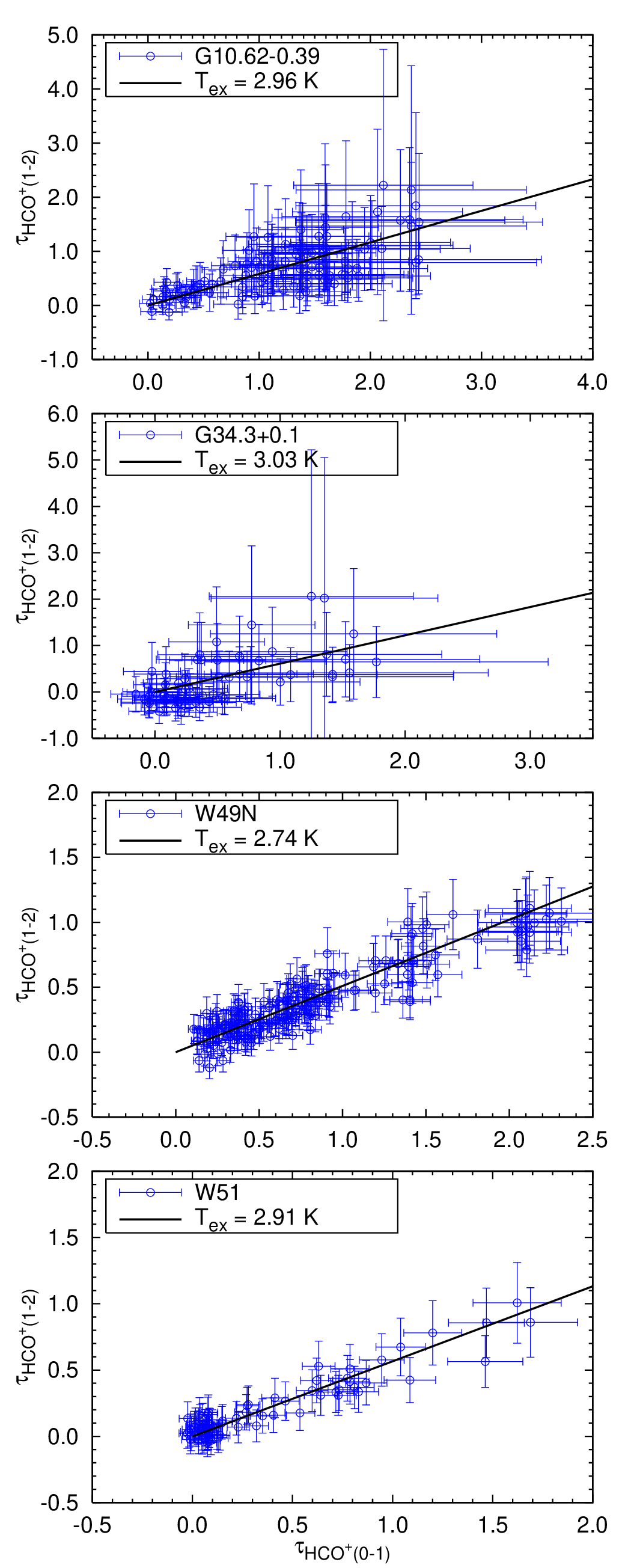}
\caption{HCO$^{+}$ (1-2) opacities as a function of HCO$^{+}$ (0-1)
  opacities. The lines are the results of linear regressions whose only
  parameter is the excitation temperature $T_{\rm ex}$.}
\label{FigComparTau}
\end{center}
\end{figure}

We observed the HCO$^{+}$ (1-2) line toward 4 of the 6
sources listed in Table \ref{TabSources}. As shown in
Fig. \ref{FigComparTau}, which displays the opacities $\tau_{{\rm HCO}^{+}
  (0-1)}$ and $\tau_{{\rm HCO}^{+} (1-2)}$ in the absorption lines channel by
channel, the rms noise levels are low enough to compute accurate values of the
excitation temperature toward W49N and W51. Toward G10.62-0.39 and 
 G34.3+0.1, the correlation between $\tau_{{\rm HCO}^{+}
  (0-1)}$ and $\tau_{{\rm HCO}^{+} (1-2)}$ is looser and 
could be due to actual variations
of the excitation temperature along the line of sight. 

However, for the sake of simplicity, we assume that the absorbing gas on each
line of sight is defined by a single excitation temperature $T_{\rm ex}$,
derived as 
\begin{equation}
\frac{1-{\rm exp}\left( -h \nu_{2} / k T_{\rm ex}\right)}{{\rm exp}\left( h
  \nu_{1} / k T_{\rm ex}\right)-1} = \frac{\int \tau_{2} d\upsilon}{\int
  \tau_{1} 
  d\upsilon} \left( \frac{\nu_{2}}{\nu_{1}} \right)^{3}
  \frac{g_{u1}A_{ul1}}{g_{u2}A_{ul2}},
\end{equation}
where $\tau_{1}$ and $\tau_{2}$ are the line opacities per unit velocity,
$\nu_{1}$ and $\nu_{2}$
the rest frequencies, $g_{u1}$ and $g_{u2}$ the upper level degeneracies, and
$A_{ul1}$ and $A_{ul2}$ the Einstein's spontaneous emission coefficients of
the molecular transitions HCO$^{+}$ (0-1) and (1-2) respectively.  The results
are listed in Table \ref{TabTex}. For all sources, $T_{\rm ex}$ is found close
to the temperature of the cosmic microwave background $T_{\rm CMB} = 2.73$ K.

\begin{table}[!h]
\begin{center}
\caption{Excitation temperatures inferred from the $\tau_{{\rm HCO}^{+}
    (1-2)}/\tau_{{\rm HCO}^{+} (0-1)}$ ratio.}
\begin{tabular}{l c c c}
\hline
Source      & \multicolumn{1}{c}{T$_{\rm ex}$ (K)} &
\multicolumn{2}{c}{errors$^{a}$ (K)}\\ 
\hline
            &       &   +  &   -    \\
G10.62-0.38 &  3.0  & 1.4  &  0.5  \\
G34.3+0.1   &  3.0  & 8.4  &  0.8  \\
W49N        &  2.7  & 0.8  &  0.4  \\
W51         &  2.9  & 1.1  &  0.5  \\
\hline
\end{tabular}
\begin{list}{}{}
\item[$^{a}$] The errors are derived from the results of numerical calculation
  which do not take into account the lower limit of 2.73 K set by the cosmic
  microwave background.
\end{list}
\label{TabTex}
\end{center}
\end{table}

These low excitation temperatures suggest that HCO$^{+}$, HNC, and HCN, whose
dipole moments are similar, are radiatively rather than collisionally
excited. Using the Large Velocity Gradients (LVG) code by Schilke (private
communication), we obtain an upper limit on the gas density: $n_{\rm H} <
10^{4}$ cm$^{-3}$. This constraint is not stringent because of the large
critical densities of the transitions
($\sim 10^{5}$ and $\sim 10^{6}$
\cc\ for HCO$^{+}$ (1-0) and HNC and HCN (1-0) respectively).

Estimates of the gas densities are  provided  by the LVG analysis of emission lines 
observed at the velocities of the absorption components.
Toward W49N, \twCO\ and \thCO(1-0) and (2-1) line observations  
provide H$_2$ densities all close to  $5 \times 10^{3}$ cm$^{-3}$ (Vastel \etal 2000).
Including the CI $^3P_1 - ^3P_0$ line in the analysis of the CO lines, Plume
\etal (2004) find lower densities, ranging between 1500 and 3000 \cc.
In the following, we therefore adopt $n_{\rm H} < 5\times 10^{3}$ \cc\ as 
an upper limit of the gas density causing the absorption features.


\subsection{Decomposition of the spectra into Gaussian components} \label{Decom-Gauss}

The decomposition of the spectra in velocity components and the resulting
column densities are inferred from a
multi-Gaussian fitting procedure based on the Levenberg-Marquardt
algorithm and the following sequence.
\begin{itemize}
\item[$\bullet$] One of the absorption spectra is decomposed without a priori into
  the minimal number of Gaussians required to fit the data within the
  observational errors. We preferentially use the CN (0-1) transition because
  its hyperfine structure provides a valuable constraint on the line centroids.
\item[$\bullet$] Based on the results of the previous step, the algorithm is
  then applied recursively to the others absorption spectra, ruling out a
  shift by more than a resolution element between the velocity components of
  each molecular species. In a few cases, a shift of at most 4 ($\sim$ 0.5
  \kms) resolution elements is allowed because it ensures the convergence of
  the fitting routine: for example in regions of the spectra where many
  components are blended, as in the HCN (0-1) data at 59.5 \kms\ toward
  W49N.
\end{itemize}
Thus, for each transition, the observed line profile (line/continuum) is
written:
\begin{equation}
  \frac{T_{\rm mb}}{T_{\rm c}}(\upsilon) = {\rm exp}\left[ - \sum_{j=1}^{N_{c}}
  \sum_{k=1}^{N_{h}} \alpha_{h}(k) \, \tau_0(j) \, e^{-\frac{1}{2}\left[\frac{\upsilon -
  \upsilon_{0}(j) - \upsilon_{0h}(k)}{\sigma_{\upsilon}(j)}\right]^{2}}\right]
\end{equation}
where $T_{\rm mb}$ and $T_{\rm c}$ are the main beam and the continuum
temperatures, $N_c$ and $N_h$ are the numbers of components and hyperfine
transitions\footnote{$N_h$ depends on the molecule}, $\tau_0$, $\upsilon_0$
and $\sigma_{\upsilon}$ are the Gaussian parameters, and $\alpha_{h}$ and 
$\upsilon_{0h}$ are the LTE and optically thin line strengths relative to a
reference hyperfine transition and the velocity shift associated to each
hyperfine transition. As described in Appendix \ref{AppendGauss}, the column
density associated to each velocity component is then derived assuming a
single excitation temperature $T_{\rm ex}$ for all the levels of a given
molecule.

\begin{figure}
\begin{center}
\includegraphics[width=6.4cm,angle=-90]{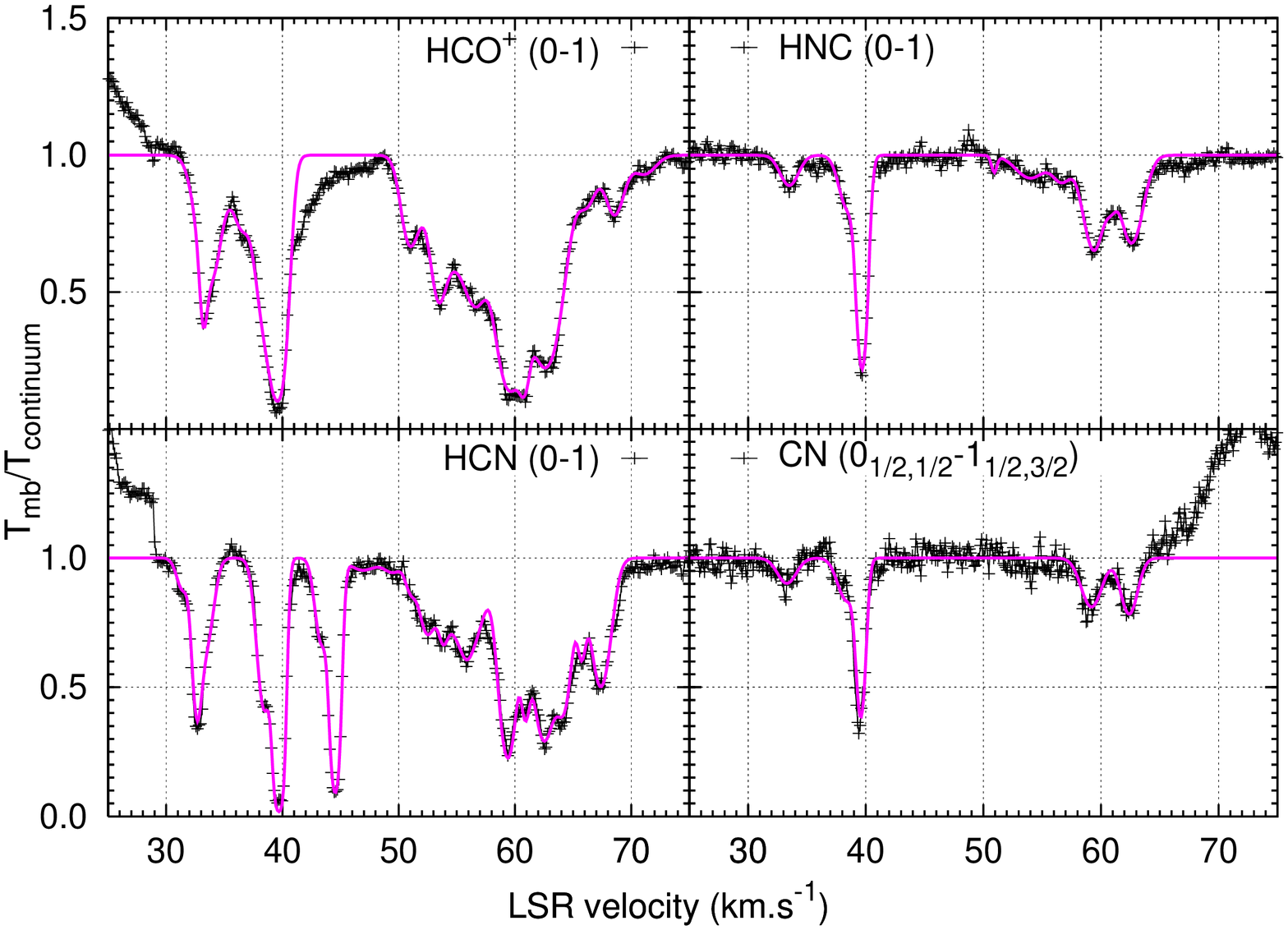}
\caption{Observational data (black points) compared to the multi-Gaussian
  decomposition (thick purple line) of the HCO$^{+}$, HNC, HCN, and CN (0-1)
  absorption spectra observed toward W49N.}
\label{FigMultigauss}
\end{center}
\end{figure}

For the components which are either saturated or too weak to be singled out by
the fitting procedure, limits on the column densities are derived from the
integrated optical depth over the corresponding velocity range. All the results
are listed in Tabs. \ref{TabFitHCOp}-\ref{TabFitCN} of Appendix
\ref{AppendGauss} and, as an example, the outcome of the multi-Gaussian
decomposition, applied to the absorption lines observed toward W49N, is shown in
Fig. \ref{FigMultigauss}. We find that the widths of the Gaussian components
 have a continuous distribution with values ranging from 0.3 \kms\ to 3.5
 \kms, independently of the source or the molecular species. The peak optical
 depths range between 0.06 and 2.2 and the inferred column densities per
 velocity component span more than one order of magnitude on each line of
 sight.

\begin{figure*}[!ht]
\begin{center}
\includegraphics[width=12.0cm,angle=0]{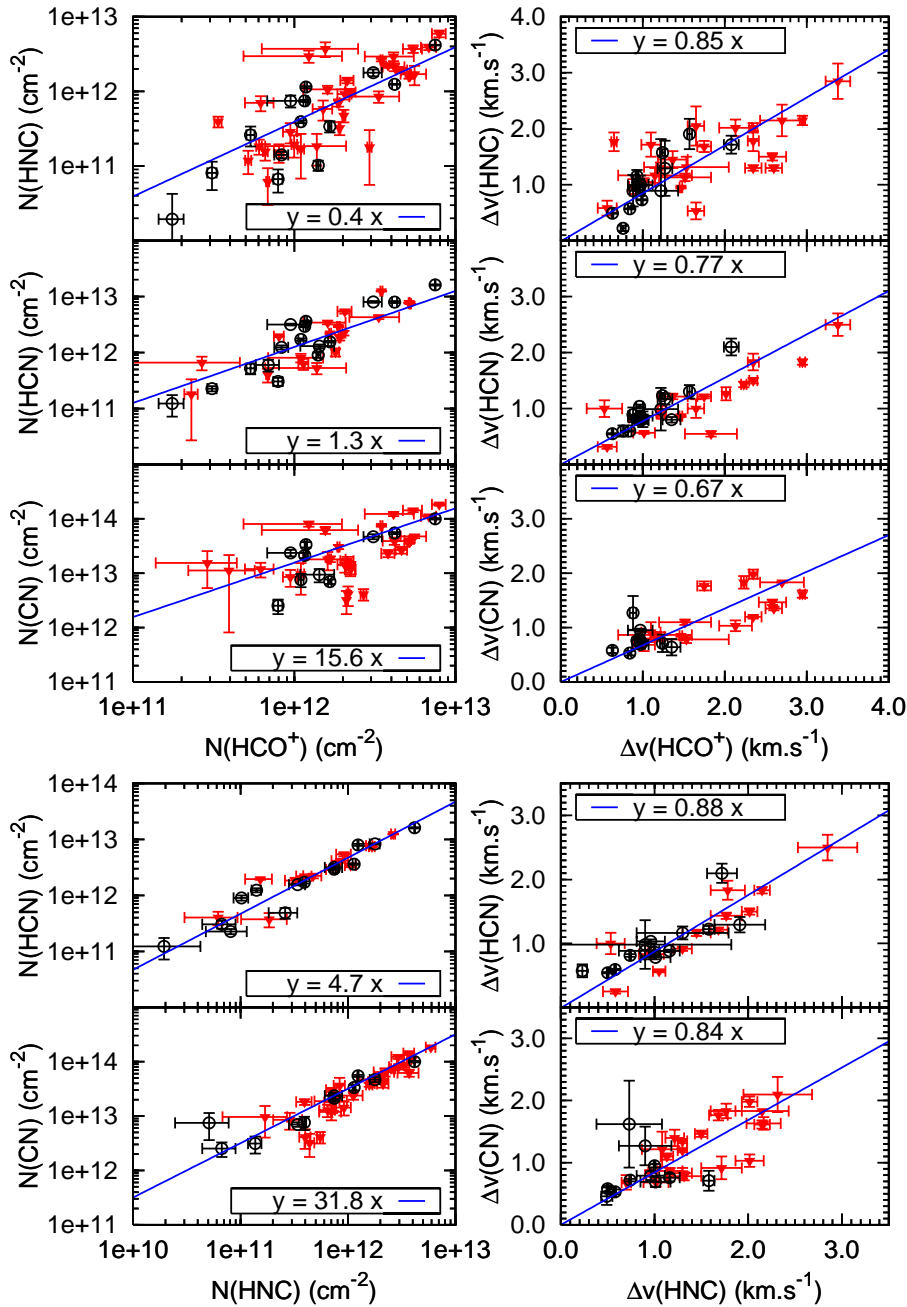}
\caption{Comparisons of the column densities (left panels) and the FWHM
  (rigth panels) of HCO$^{+}$, CN, HCN, and HNC. The red filled symbols are
  from this work. The black open symbols are from Lucas \& Liszt (1996) and
  Liszt \& Lucas (2001). The solid lines result from a linear regression of
  the data of the present work
  unweighted by the fractional errors on $N$ and $\Delta \upsilon$ : Y =
  $\alpha$ X.}
\label{FigCorrelation1}
\end{center}
\end{figure*}

\subsection{Systematic errors on calculation of column densities} \label{SectErrors}

The uncertainties given in
Tabs. \ref{TabFitHCOp}-\ref{TabFitCN} are the formal 1-$\sigma$ errors derived
from the diagonal elements of the covariance matrix and 
do not take into
account the systematic  errors introduced by: (1) the finite velocity
resolution, (2) the uncertainty on the excitation temperatures (see
Table \ref{TabTex}), and (3) the uncertainty on the continuum level.

We show in Appendix \ref{AppendAbscissa} that the finite velocity resolution
introduces an error on the column densities smaller than 12 \%.
Table \ref{TabTex} gives the uncertainties on the excitation
temperature of the lowest rotational levels of HCO$^{+}$, that affects the
partition function, hence the column densities. Toward G10.62-0.38, W49N, and
W51, the uncertainties on $T_{\rm ex}$ are small, providing uncertainties on
the column densities $<^{+50\%}_{-10\%}$. Toward G34.3+0.1, the inferred
column densities could be underestimated by a factor of 3.

Finally, the uncertainty $\epsilon = \delta T_{\rm c}/T_{\rm c}$ on the
continuum temperature (see Sect. 2) introduces an error $\delta \tau$ on the
calculation of the optical depths:
\begin{equation}
\delta \tau \sim \epsilon -{\rm ln} \left( 1 + \epsilon e^{\tau} \right).
\end{equation}
This error is larger, in most cases, than those inferred from the fitting
 procedure, and for $\epsilon = 10$\% the corresponding uncertainty on the
 column densities ranges from 16\% to 62\% when $\tau_0$ varies between 0.5
 and 2. All components with $\tau_0 > 2.2$ are thus considered as
 saturated. The components at 16 and 41 \kms\ of the G10.62-0.38
 HCO$^{+}$(0-1) spectrum are the only ones that fall in this category.

When all uncertainties are taken into account, the column densities are
determined within a factor of 2, and within a factor of 3 for the absorption
lines observed toward G34.3+0.1.

\section{Results} \label{SectRes}

\subsection{Comparison of column densities and profile linewidths among species}

\begin{figure}[!h]
\begin{center}
\includegraphics[width=8.0cm,angle=0]{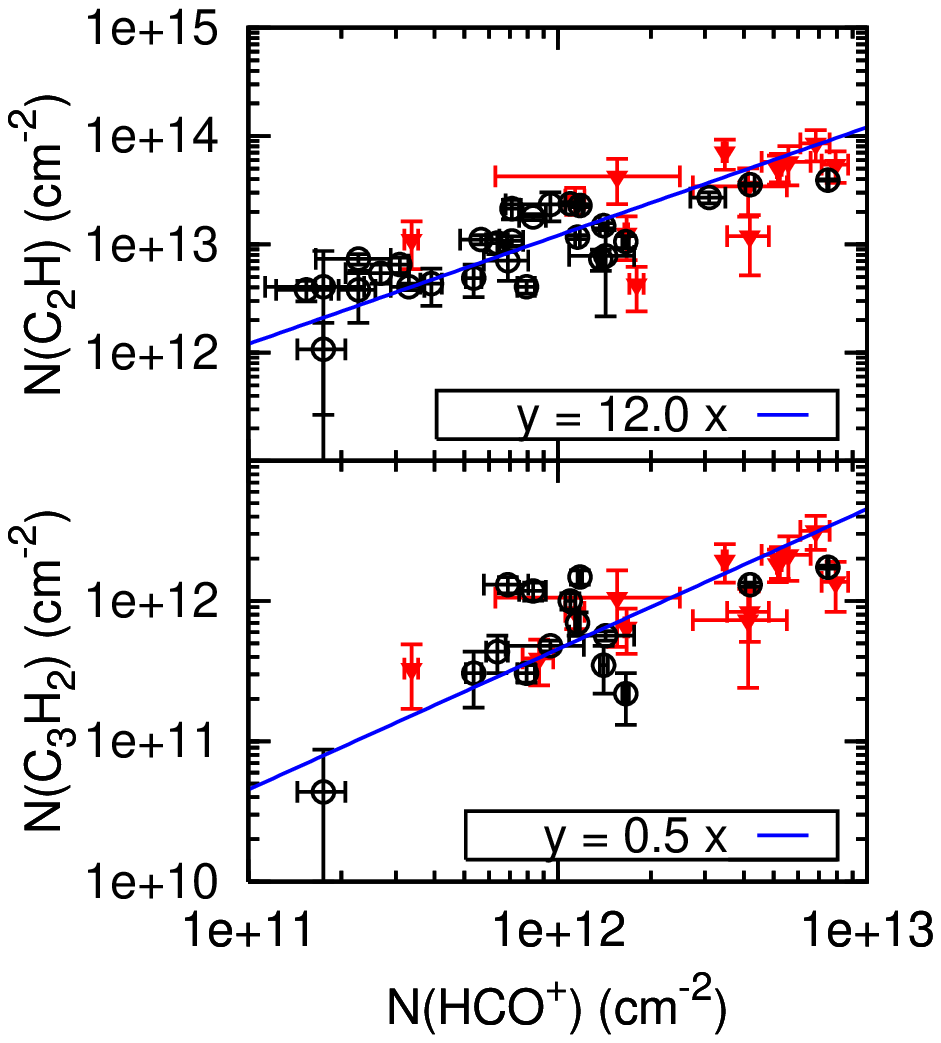}
\caption{Comparisons of the column densities of HCO$^{+}$, C$_2$H, and
  $c-$C$_3$H$_2$. The column densities of C$_2$H and $c-$C$_3$H$_2$ are from
  Gerin \etal (2010). The red filled symbols are from this work. The black
  open symbols are from Lucas \& Liszt (1996,2000). The solid lines result
  from a linear regression  of the data of the present work 
  unweighted by the fractional errors on $N$ : Y =
  $\alpha$ X.}
\label{FigCorrelation2}
\end{center}
\end{figure}

Because the line centroids of a Gaussian component observed in several
transitions coincide within 0.15 km s$^{-1}$ (except for 6
  components, see Appendix \ref{AppendGauss},
  Tabs. \ref{TabFitHCOp} - \ref{TabFitCN}), we propose that the
corresponding velocity component has a physical reality, in support of the
comparison between the optical depths in the different molecular lines and the
resulting column densities.

The column densities and linewidths of all the unsaturated velocity components
are displayed in Figs. \ref{FigCorrelation1} and \ref{FigCorrelation2}
for several pairs of species. These figures also include, for comparison, the
results of previous studies of molecular absorption lines observed toward
strong extragalactic mm-wave continuum sources (Lucas \& Liszt 1996, 2000 ;
Liszt \& Lucas 2001) that mostly sample diffuse gas in the Solar
Neighbourhood.
 
Although we have not made any assumption on the nature of a Gaussian velocity
component, it is remarkable that {\it (i)} the range of the component
linewidths (0.3 to 3.5 \kms) is about the same in both data sets, and {\it
  (ii)} the range  of column densities of most  molecular species  (up to two
orders of magnitude) is also similar in both sets. This result suggests that
the gas components sampled  at high galactic latitude  and toward the inner
Galaxy share common kinematic properties. In addition the rotation curve
in the first quadrant provides displacements  $\sim 70 - 100$ pc/km s$^{-1}$:
therefore, if we assume that the induced spreading of the velocity centroids
is responsible for the observed linewidths, a component of 3 km s$^{-1}$
(resp. 0.3 km s$^{-1}$) would have a size of 300 pc (resp. 30
pc). This would correspond to densities $n_H < 40$ cm$^{-3}$, below our
estimates (see Sect. \ref{PhysPropGas}). 
 Hence, the   linewidths of the components are likely due to turbulence
 rather than differential  rotation of the galactic plane.
In the following, we compare the chemical and kinematic
properties  of these two ensembles. While the thick lines in
Figs. \ref{FigCorrelation1}  and \ref{FigCorrelation2} result from a linear
regression unweighted by the fractional errors on the fitting parameters, the
values given below come from weighted\footnote{A weighted average
  makes more sense numerically but tends to favour the high column density
  points. In practice, both evaluations (weighted and unweighted) are
  meaningful since the errors are not known with high accuracy (see
  Sect. \ref{SectErrors}).} averages.

\begin{itemize}
\item[$\bullet$] Concerning the column densities of HNC, HCN, and CN, Liszt \&
  Lucas (2001) found tight linear correlations along lines of sight at high
  Galactic latitude, namely 
$N({\rm HCN}) / N({\rm HNC}) = 4.8 \pm 1.1$ and
$N({\rm CN }) / N({\rm HNC}) = 33 \pm 12$.
In comparison the mean ratios of the present work are
$N({\rm HCN}) / N({\rm HNC}) = 4.8 \pm 1.4$ and
$N({\rm CN }) / N({\rm HNC}) = 34 \pm 12$.

\item[$\bullet$] Fig. \ref{FigCorrelation1} (top)  shows
that the correlations of the CN-bearing molecules with \HCOp\ are looser   
and probably non linear with mean ratios:   
$N({\rm HNC}) / N({\rm HCO}^{+}) = 0.5 \pm 0.3$ ,
$N({\rm HCN}) / N({\rm HCO}^{+}) = 1.9 \pm 0.9$, and
$N({\rm CN }) / N({\rm HCO}^{+}) = 18  \pm 9$.

\item[$\bullet$] Last, the loose linear relation between the column densities
  of \CtH\ and \HCOp\  observed in Fig. \ref{FigCorrelation2} corresponds to  a
  mean ratio $N({\rm C}_2{\rm H}) / N({\rm HCO}^{+}) = 8.3 \pm 5.6$ 
  to be compared with $N({\rm C}_2{\rm H}) / N({\rm HCO}^{+}) = 14.5 \pm 6.7$
  obtained by Liszt \& Lucas (2001).
\end{itemize}

The linewidths of the Gaussian components of the HCO$^{+}$, CN, HCN, and HNC
 line profiles are close to be the same. We find that the CN, HCN, and HNC line
 profiles are systematically narrower than those of HCO$^{+}$ (by a factor of
 0.7 to 0.9). In Fig. \ref{FigCorrelation1}, the CN and HCN line profiles
 appear to be narrower than those of HNC (by a factor of 0.8 to 0.9) but this
 result is mostly due to the hyperfine structure of HNC that has not been
 taken into account  in the decomposition. Once the  correction discussed in
 Sect. \ref{SectObs} is applied, the significance of the linewidth difference
 disappears. The profiles of  C$_2$H and C$_3$H$_2$ have been studied with a
 different method,  and the components identified are not the same as those in
 the present study (Gerin \etal 2010).

\subsection{Estimation of the total hydrogen column densities along the lines
 of sight} \label{EstimationOfHydrogen}

In order to be able to compare the data to chemical models, molecular
abundances must be  derived and column densities of hydrogen measured. In
particular, it is essential  to determine whether the wide dynamic ranges
over which the correlations are observed (Figs. \ref{FigCorrelation1} and
\ref{FigCorrelation2}) are related to variations of the total column density
of the gas sampled or/and of the physical and chemical conditions in the
absorbing gas.
The main difficulty is to estimate the fraction of molecular hydrogen that is 
not directly observable. Using $\lambda 21$ cm observations of HI, $\lambda
9$ cm and $\lambda 0.3$ mm observations of CH, and the remarkable correlation
between CH and H$_2$,
$N({\rm CH})/N({\rm H}_2) = 4.3 \times 10^{-8}$ (Liszt \& Lucas
2002), we evaluate the total amount of gas $N_{\rm H} =
N({\rm H}) + 2N({\rm H}_2)$ along the galactic lines of sight, 
 as in Godard \etal (2009) for the lines of sight studied by Liszt \& Lucas
(2001). The HI column densities are inferred from
VLA $\lambda 21$ cm absorption line observations (Koo \etal 1997; 
Fish \etal 2003). Wherever possible, we derive $N({\rm H}_2)$
from CH observations (at 9 cm by Rydbeck \etal 1976, at 1 THz by Gerin \etal
2010 in prep.).

An independent estimate of the total column density of gas 
toward the star-forming regions is inferred from the analysis of the 2MASS
survey  (Cutie \etal 2003). Marshall
\etal (2006) have measured the near infrared colour excess in large areas of 
the inner Galaxy ($|l|<100^{\circ}$, $|b|<10^{\circ}$) to obtain the
visible extinctions ($A_V \sim 10 A_K$), providing an estimate
of the total hydrogen column density along the lines of
sight.

\begin{table}[!h]
\begin{center}
\caption{HI, H$_2$ and total hydrogen column densities.}
\begin{tabular}{l @{\hspace{0.3cm}} r @{\hspace{0.3cm}} r @{\hspace{0.3cm}}
 r @{\hspace{0.3cm}} r @{\hspace{0.3cm}} r}
\hline
            & $\upsilon_0$         & $\Delta \upsilon$    & $N$(H)$^{a}$ & $N({\rm H}_2)^{c}$ & $N_{\rm H}^{d}$ \\
Source      & (\kms) & (\kms) & (cm$^{-2}$) & (cm$^{-2}$) & (cm$^{-2}$) \\
            &               &               & $\times 10^{21}$ & $\times 10^{21}$ & $\times 10^{21}$ \\
\hline
{\bf W51}   &
$\left. \begin{array}{r r r}   6.2 \\  11.8 \end{array} \right. $ & 
$\left. \begin{array}{r r r}   5.6 \\   6.0 \end{array} \right. $ &
$\left. \begin{array}{r r r}  1.27 \\  0.58 \end{array} \right. $ &
0.47 & 2.3 \\
\\
\hline
            & $\upsilon_{\rm min}$ & $\upsilon_{\rm max}$ & $N$(H)$^{b}$ & $N({\rm H}_2)^{c}$ & $N_{\rm H}^{d}$ \\
Source      & (\kms) & (\kms) & (cm$^{-2}$) & (cm$^{-2}$) & (cm$^{-2}$) \\
            &               &               & $\times 10^{21}$ & $\times 10^{21}$ & $\times 10^{21}$ \\
\hline
{\bf G05.88-0.39} & 
$\left. \begin{array}{r r r} -31.3 \\ -10.0 \\  4.0 \\  20.0 \end{array} \right. $ & 
$\left. \begin{array}{r r r} -10.0 \\   4.0 \\ 20.0 \\  29.3 \end{array} \right. $ &
$\left. \begin{array}{r r r}  1.59 \\ >6.68 \\ 5.87 \\ >3.86 \end{array} \right. $ &
 & 16 \\
\\
{\bf G08.67-0.36} & 
$\left. \begin{array}{r r r}  -1.7 \\  11.3 \\  25.7 \end{array} \right. $ & 
$\left. \begin{array}{r r r}  11.3 \\  25.7 \\  45.3 \end{array} \right. $ &
$\left. \begin{array}{r r r}  4.80 \\  5.32 \\ >9.62 \end{array} \right. $ &
 & 14 \\
\\
{\bf G10.62-0.38} &
$\left. \begin{array}{r r r}  11.3 \\  25.0 \\ 30.6 \end{array} \right. $ & 
$\left. \begin{array}{r r r}  25.0 \\  30.6 \\ 47.3 \end{array} \right. $ &
$\left. \begin{array}{r r r}  3.53 \\  2.61 \\ 5.54 \end{array} \right. $ &
15.8 & 17 \\
\\
{\bf G34.3+0.1}   &
$\left. \begin{array}{r r r}  -2.5 \\   7.1 \\ 20.0 \end{array} \right. $ & 
$\left. \begin{array}{r r r}   7.1 \\  20.0 \\ 34.1 \end{array} \right. $ &
$\left. \begin{array}{r r r}  0.54 \\  2.01 \\ 4.42 \end{array} \right. $ &
 & 9 \\
\\
{\bf W49N}        &
$\left. \begin{array}{r r r}  30.0 \\  50.0 \end{array} \right. $ & 
$\left. \begin{array}{r r r}  50.0 \\  78.2 \end{array} \right. $ &
$\left. \begin{array}{r r r}  6.95 \\  7.23 \end{array} \right. $ &
$\left. \begin{array}{r r r}   1.6 \\   4.0 \end{array} \right. $ &
23 \\
\hline
\end{tabular}
\begin{list}{}{}
\item[$^{a}$] Gaussian decomposition from Koo (1997). The column densities
  have been scaled to a spin temperature of 100 K.
\item[$^{b}$] Inferred from the absorption profiles observed by Fish et
  al. (2003) with the VLA interferometer. A spin temperature of 100 K is
  assumed.
\item[$^{c}$] Estimation from the CH emission lines observed by Rydbeck et
  al. (1976) toward W51 and W49N, the CH absorption lines observed by Gerin
  \etal (2010 in prep.) toward G10.62-0.38, and the correlation between CH and H$_2$,
  $N({\rm CH})/N({\rm H}_2) = 4.3 \times 10^{-8}$ (Liszt \& Lucas 2002).
\item[$^{d}$] From models of the extinction at 2 $\mu$m by Marshall et
  al. (2006).
\end{list}
\label{TabHI}
\end{center}
\end{table}

Table \ref{TabHI} lists the HI (and H$_2$, where available) 
column densities in selected velocity intervals, as well 
as the total hydrogen column densities inferred from extinction.
However the uncertainties on these estimations are large: (1) the error
  on the $N({\rm CH})-N({\rm H}_2)$ relation is about a factor of 3 (Liszt \&
  Lucas 2002) ; (2) this correlation has been established in the local diffuse
  medium but has never been observed in the inner Galaxy material ; (3)
  because of the low resolution of the 2MASS survey ($\sim 15$
  arcmin), the  error on the total hydrogen column density (computed as the
  standard deviation of the extinction measured along the four closest lines
  of sight surrounding a given source) is
  larger than 30 \% ; and (4) The HI column densities inferred from VLA
  $\lambda 21$ cm absorption line observations are directly proportional to
  the assumed spin temperature.
  Hence, while the two determinations agree with each other
  within 15\% toward W51 and W49N, they differ by at least a factor of 2
  toward G10.62-0.38.

According to the extinction measurements, the lines of sight sample between 1.3
(W51) and 12.5 (W49N) magnitudes of gas.
The total velocity coverage of the absorption features is $\sim 10$ \kms\
toward W51 and 48 \kms\ toward W49N. Therefore, the average hydrogen column
density {\it per velocity unit}  is only twice larger along the line of sight
toward W49N than along that toward W51. 
It is therefore possible to estimate the molecular abundances relative to the
total hydrogen column density $N_{\rm H}$
for each velocity component, assuming that $N_{\rm H}$ scales with 
their linewidth according to $N_{\rm H}/\Delta v = 2.2$ and 4.7$\times
10^{20}$  cm$^{-2}$/\kms\ toward W51 and W49N respectively. 
This is equivalent to assuming a uniform HI optical depth 
in the gas components where we observe molecular absorption. Such an
approximation  underestimates the  HI column density per unit velocity by no
more than a factor 2. The total hydrogen column densities estimated with this
method and given in Table A.1 for W49N and W51  are smaller by only 50\% than
those inferred from IR extinction. We therefore estimate that the total H
column density per velocity component on these two  lines of sight does not
exceed 1.5 magnitude (or about $2.5 \times 10^{21}$ cm$^{-2}$).
This result suggests that the \HCOp\ abundance in the different velocity
components, defined as 
$X({\rm HCO}^{+}) = N({\rm HCO}^{+})/N_{\rm H}$,
ranges between 5$\times 10^{-10}$ and 
5$\times 10^{-9}$ toward both W49N and W51. For HCN, the scatter of abundances
is also  an order of magnitude among the components, with values ranging
between  2$\times 10^{-9}$ and 3$\times 10^{-8}$ for W49N. A similar scatter
is also found for HNC with  abundances $\sim 10$ times smaller.

In summary, since the column densities of the Gaussian components span less
than two orders of magnitude while their linewidths span only a factor $\sim
10$ (between 0.3 and 3.4 \kms),  actual fluctuations  of molecular abundances
are therefore observed among the components.

\subsection{Physical properties of the gas seen in absorption} \label{PhysPropGas}

The above average description is crude and is only meant to ascribe an
  average column density to a velocity interval, ignoring velocity crowdings:
  the lines of sight toward distant star-forming regions sample gas components
  with a broad distribution of densities, velocity dispersions, and column
  densities.
We have shown that the upper limit on the gas density
is $n_{\rm H} < 5 \times 10^{3}$ \cc\ (cf. Sect. \ref{SectTemp})
and that the total column densities per velocity components $N_{\rm H}$ are at
most of the order of a few magnitudes, similar to those obtained along
the high latitude lines of sight observed by Liszt \& Lucas (2001).
According to the  definitions of  Snow \& McCall (2006), the gas sampled by
these lines of sight is a mixture  of diffuse ($n_{\rm H} < 500$ cm$^{-3}$
with a shielding from the UV field $A_V < 1$) and translucent gas ( $500$
cm$^{-3}$ $< n_{\rm H} < 5000$ cm$^{-3}$ and a shielding $1 < A_V < 2$).

Allowing the gas density to range between 30 cm$^{-3}$ and $5\times
  10^{3}$ cm$^{-3}$, the total N$_{\rm H}$ per velocity component of a few
  magnitudes translates into sizescales ranging between $\sim 0.2$ and $\sim
  30$ pc. It is interesting that the corresponding range of velocity
  dispersions inferred from the linewidth-scale relation in the diffuse
  molecular gas (Falgarone \etal 2009) is $\sim 0.2 - 5$ km s$^{-1}$, very
  close to the observed range of component velocity dispersions. It suggests
  that the mixture of diffuse and transluscent gas sampled by these lines of
  sight is entrained in the turbulent interstellar cascade. Although one
  velocity component is not necessarily a cloud, the linewidth-scale relation
  ascribes a sizescale to each velocity dispersion, $\Delta u \sim 1 (l/1\,\,
  {\rm pc})^{1/2}$ km s$^{-1}$. Therefore, different molecular species present
  in a given velocity component of width $\Delta u$ are colocated within $\sim
  0.1$ pc (if $\Delta u = 0.3$ km s$^{-1}$) and 9 pc (if $\Delta u = 3$ km
  s$^{-1}$). It is the correlation of molecular species, observed at these
  grains, that we will now confront to the predictions of different numerical
  codes.

In the following sections, we therefore study the chemistry of diffuse and
translucent gas, and we compare the observed column densities to the results
of PDR (PhotoDissociation Region) and TDR (Turbulent Dissipation Region)
models.

\section{Comparison of observations to model predictions} \label{SectDis}

\begin{figure*}[!!ht]
\begin{center}
\includegraphics[width=12.0cm,angle=0]{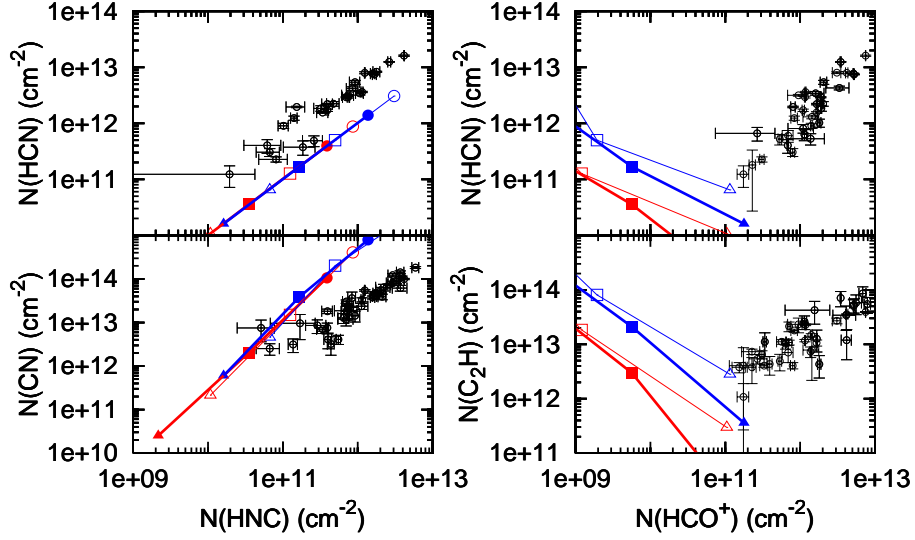}
\caption{Observations compared to the predictions of PDR models. The data
  (open circles) are from Lucas \& Liszt (2000), Liszt \& Lucas (2001), Gerin
  \etal (2010), and this work. The PDR models are computed
  for several densities:  $10^{2}$ (triangles), $10^{3}$ (squares), and
  $10^{4}$ (circles) cm$^{-3}$. Red and blue curves correspond to $N_{\rm H}=
  1$ and 2 magnitudes respectively. Thin (empty symbols) and thick
  (filled symbols) curves correspond to $\chi = 1$ and $\chi = 3$.}
\label{FigPDR}
\end{center}
\end{figure*}

\subsection{Introduction to the PDR and TDR models}

The comparison of the PDR and TDR models has been discussed in detail in
Godard \etal (2009). We recall here the main differences between the two
approaches.

The PDR model is a 1-dimensional chemical model in which 
a static slab of gas of uniform density and given thickness (or total column
density of gas, noted $N_{\rm H}$ in the following) 
is illuminated by the ambient interstellar radiation field
either on one side or on both sides (Le Petit \etal 2006). 
The computed column density of a molecular species to be compared to the 
observed values  is therefore an integral performed over the slab thickness
along a direction perpendicular to its surface.

The TDR code is a 1-dimensional model in which the chemical and thermal
evolution of a turbulent dissipative burst - namely a magnetized vortex - is
computed. The dynamics and lifetime of the magnetized vortices are controlled
by the
turbulent rate of strain $a$ (in s$^{-1}$), the gas density $n_{\rm H}$ (in
cm$^{-3}$), and the maximal orthoradial velocity $u_{\theta \,{\rm max}}$ (in
cm s$^{-1}$). A random line of sight crossing several dissipative regions is
then modelled by taking into account: the averaged turbulent dissipative
rate observed in the CNM, and the long-lasting thermal and chemical
relaxation stage that follows any dissipative burst. The resulting line of
sight therefore intercepts three kinds of diffuse gas: (1) mainly the ambient 
medium (the corresponding filling factor is larger than 90 \%) in which
the chemistry is driven by the UV radiation field, (2) the active vortices
where the gas is heated and the chemistry is enhanced by the dissipation of
turbulent energy, and (3) the relaxation stages where the gas previously
heated cools down to its original state.
Because of the strong experimental (Mouri \etal 2007, Mouri \& Hori 2009) and
observational (Crovisier 1981 ; Joncas \etal 1992 ; Miville-Desch\^enes \etal
2003, Haud \& Kalberla 2007) constraints on $u_{\theta \,{\rm max}}$, only two
main parameters govern the TDR model: the turbulent rate of strain that
describes the coupling between the large and small scales of turbulence and
the gas density. However, both parameters are not independent because of the
  constraint imposed by the observed turbulent energy available in the CNM,
  namely ${\overline \varepsilon}_{obs} \sim 2 \times 10^{-25}$ \eccs
  (Falgarone 1998, Hily-Blant \etal 2008). 
As a result, the influence of the parameters on the TDR model, fully discussed
  in Godard \etal (2009), is complex.
\begin{itemize}
\item[$\bullet$] The gas density $n_{\rm H}$ affects the chemistry itself and
  has several impacts due to the dynamics and the geometry of
  the line of sight. For the range of parameters explored here, the 
  dissipation of turbulent energy over one vortex is mainly due to ion-neutral
  friction and varies roughly as $n_{\rm H}^{2}$ ; the vortex radius varies as
  $n_{\rm H}^{-1/2}$, and their number along the line of sight as $n_{\rm
  H}^{-2}$.
\item[$\bullet$] While the dissipation of turbulent energy over one vortex and
  the number of vortices along the line of sight are nearly independent of the
  turbulent rate of strain $a$, their size and their lifetime vary roughly
  as $a^{-1/2}$ and $a$ respectively.
\end{itemize}

For both models and as shown in Table \ref{TabChem}
of Appendix \ref{AppendChemistry}, the nitrogen gas phase chemistry has been
updated according to the recent measurements and rate review. However,
neither the PDR nor the TDR chemical networks take into account the nitrogen
grain surface chemistry: this assumption will be discussed in
Sect. \ref{GrainChemistry}.

\subsection{Gas phase nitrogen chemical networks in the PDR and TDR models}

The networks of dominant reactions in the nitrogen gas phase chemistry
  are shown in Figs. \ref{FigChemistry0} \& \ref{FigChemistry} (Appendix
\ref{AppendChemistry}). While in both models, photodissociation largely
  dominates the destruction mechanisms of CN-bearing molecules, there are
  several  pathways that lead to their production.

 In the PDR model (Federman \etal 1994, Boger \& Sternberg 2005), two main
formation routes are at work. The first
involves the hydrogenation chain of carbon and the formation of CH and
CH$_2$, which in turn, leads to CN and HCN via the neutral-neutral reactions:
\begin{equation}
{\rm CH} + {\rm N} \rightarrow {\rm CN} + {\rm H}
\end{equation}  
and
\begin{equation}
{\rm CH}_2 + {\rm N} \rightarrow {\rm HCN} + {\rm H}.
\end{equation}
The second one involves the hydrogenation chain of nitrogen and the formation
of NH and NH$_2$, followed by the ion-neutral reaction chains
\begin{equation}
{\rm NH}     \overset{{\rm C}^{+}  }{\longrightarrow} 
{\rm CN}^{+} \overset{{\rm H}_2    }{\longrightarrow} 
\left.
\begin{array}{l}
{\rm HCN}^{+} \\
{\rm HNC}^{+}
\end{array}
\right.
\overset{{\rm H}_2}{\longrightarrow}
{\rm HCNH}^{+}
\end{equation}  
and
\begin{equation}
{\rm NH}_2    \overset{{\rm C}^{+}  }{\longrightarrow} 
{\rm HCN}^{+} \overset{{\rm H}_2    }{\longrightarrow}
{\rm HCNH}^{+},
\end{equation}
the dissociative recombination of HCNH$^{+}$, and the photodissociation
  of HCN and HNC.


In the TDR model, because
of both the differential rotation at the edge of the vortex and the
ion-neutral drift, the gas is locally heated, and several  key endoenergetic
reactions such as 
\begin{equation}
{\rm C}^{+} + {\rm H}_{2} \rightarrow {\rm CH}^{+} + {\rm H}   \qquad - \Delta
E/k = 4\,640 {\rm K}
\end{equation}
open up.
A warm nitrogen chemistry is triggered locally
(Fig. \ref{FigChemistry}, Appendix \ref{AppendChemistry}). Although
the temperatures reached in a vortex ($\sim 10^{3}$ K) are not sufficient to
activate the reaction 
\begin{equation} \label{EqNpH2}
{\rm N} + {\rm H}_{2} \rightarrow {\rm NH} + {\rm H}   \qquad - \Delta E/k =
16\,606 {\rm K} 
\end{equation}
(whose rate becomes large enough only at 7000 K, Crawford \& Williams 1997), 
the pathways leading to the production of
N bearing species are deeply modified and their abundances increase
considerably. The opening of the endoenergetic route C$^{+} + $ H$_2$,
  that forms CH$^{+}$ and then CH$_2$$^{+}$ and CH$_3$$^{+}$ by fast reaction
  with H$_2$, triggers in turn
the enhancement of the production of CN-bearing molecules through the
ion-neutral reaction chains
\begin{equation} \label{EqReacChain1}
{\rm CH}^{+}  \overset{{\rm N}  }{\longrightarrow} 
{\rm CN}^{+}  \overset{{\rm H}_2}{\longrightarrow} 
\left.
\begin{array}{l}
{\rm HCN}^{+} \\
{\rm HNC}^{+}
\end{array}
\right.
\overset{{\rm H}_2}{\longrightarrow}
{\rm HCNH}^{+},
\end{equation}
\begin{equation} \label{EqReacChain2}
{\rm CH}_{2}^{+} \overset{{\rm N}  }{\longrightarrow} 
{\rm HCN}^{+}    \overset{{\rm H}_2}{\longrightarrow} 
{\rm HCNH}^{+},
\end{equation}
and
\begin{equation} \label{EqReacChain3}
{\rm CH}_{3}^{+} \overset{{\rm N}  }{\longrightarrow}
{\rm HCNH}^{+},
\end{equation}
followed by the dissociative recombination of HCNH$^{+}$ and the
photodissociation of HCN and HNC.
Therefore, the key species here is CH$_3^+$. Its reaction with atomic nitrogen 
is able to significantly enhance the production of HCNH$^+$ compared to
that in the PDR models, leading to HCN, HNC, and CN. Similarly,
CH$_3$$^{+}$ is responsible for the  enhancement of \HCOp\ by its reaction
with atomic oxygen (see the discussion in  Godard \etal 2009).

\subsection{Comparison of PDR model predictions with the observations}


To explore the role of the parameters in the Meudon PDR code (Le Petit \etal
2006)  we computed several two-sided illuminated PDR models, as follows:
\begin{itemize} 
\item[$\bullet$] based on the results of Sect. \ref{EstimationOfHydrogen}, the total
  column density of gas in the slab is either $N_{\rm H} = 1.8 \times 10^{21}$
  or $3.6 \times 10^{21}$ cm$^{-2}$ (corresponding to 1 and 2 magnitudes
  respectively);
\item[$\bullet$] Two values of $\chi$ are considered, $\chi = 1$ and $\chi=3$, in order
  to bracket the ambient UV radiation field found to vary between the Solar
  Neighbourhood ($\sim 8.5$ kpc from the Galactic centre) and the molecular ring
  ($\sim 4$ kpc from the Galactic centre) by a factor $\sim 3$ (Moskalenko \etal 2006);
\item[$\bullet$] the gas density is set to $n_{\rm H} = 10^{2}$, $10^{3}$, and
  $10^{4}$ cm$^{-3}$.
\end{itemize}
Each model was computed assuming a cosmic ray ionisation rate $\zeta = 3
\times 10^{-17}$ s$^{-1}$ (Dalgarno 2006). The results on the column densities
and the column density ratios of HCO$^{+}$, HCN, HNC, CN, and C$_2$H are given in
Fig. \ref{FigPDR} which also displays, for comparison, the ranges of observed values.

This figure shows that the constraints provided by the CN-bearing molecules
and by \HCOp\ cannot be reconciled in the framework of this
model. On the one hand, both the range of column densities and column density
ratios of CN, HCN, HNC, and C$_2$H can be explained by PDR models for $N_{\rm
  H} > 1.8 \times 10^{21}$ cm$^{-2}$ and a gas density larger than 300
cm$^{-3}$. But on the other hand, for the same range of density, not only the
predicted correlation between the column densities of CN and HNC is clearly
non-linear and mismatch the observed value by a factor larger than 10, but
the column densities of HCO$^{+}$ are also underestimated by at least one
order of magnitude.

As expected, if the mean interstellar UV radiation field increases, the
  column densities of CN, HCN, HNC, and C$_2$H decrease because the
  photodissociation is the main destruction process of all these
  molecules. Oppositely,  when $\chi$ is multiplied by 3, the temperature at
  the edge of the PDR is 
  twice larger and the O + H$^{+}$ charge exchange reaction (with an
  endothermicity $\Delta E/k$ of 227 K) is enhanced. Since this reaction is at
  the root of
  the formation of HCO$^{+}$ in low density UV-dominated gas phase chemistry
  (Godard \etal 2009) and since the main destruction mechanism of HCO$^{+}$ is
  its dissociative recombination (whose rate is independent of $\chi$), the
  column density of HCO$^{+}$ increases (see Fig. \ref{FigPDR}). All those
  behaviours holds for the ranges of density and radiative conditions explored
  in Fig. \ref{FigPDR}.

The dependence of the results on the cosmic ray ionisation rate $\zeta$ is
slightly more complex.
When $\zeta$ is multiplied by 10, the abundances of H$^{+}$ and
H$_3$$^{+}$ rise by one order of magnitude\footnote{H$^{+}$ and
    H$_3$$^{+}$ are formed via the reaction chains
$
{\rm H}_2      \overset{{\rm CRP} }{\longrightarrow} 
{\rm H}_2^{+}  \overset{{\rm H}_2 }{\longrightarrow}
{\rm H}_3^{+}
$
and
$
{\rm He}     \overset{{\rm CRP} }{\longrightarrow} 
{\rm He}^{+} \overset{{\rm H}_2 }{\longrightarrow} 
{\rm H}^{+}
$
respectively (CRP: cosmic ray particle).}. Thus, both the hydrogenation
chains of nitrogen and oxygen are stimulated by the reactions
\begin{equation}
{\rm N} + {\rm H}_3^{+} \rightarrow {\rm NH}_2^{+} + {\rm H}
\end{equation}
and
\begin{equation}
{\rm O} + {\rm H}^{+} \rightarrow {\rm O}^{+} + {\rm H}
\end{equation}
respectively, while the hydrogenation chain of carbon, initiated by the
slow radiative association 
\begin{equation}
{\rm C}^{+} + {\rm H}_2 \rightarrow {\rm CH}_2^{+} + {\rm H}
\end{equation}
is independent of $\zeta$. Therefore when $\zeta$ is multiplied by 10:
\begin{itemize}
\item the column density of HCO$^{+}$ increases by a factor of 10 whatever the
  gas density ;
\item since CN always originates from the carbon hydrogenation chain, its
  column density is independent of $\zeta$ ;
\item since HCN and HNC originate from the carbon hydrogenation chain at large
  density and from the nitrogen hydrogenation chain at low density, their
  column densities are independent of $\zeta$ if $n_{\rm H} > 300$ cm$^{-3}$,
  and increase by a factor of 2 if $n_{\rm H} < 300$ cm$^{-3}$.
\end{itemize}
Nevertheless, even at low density, the impact of the cosmic ray ionization
rate is not sufficient enough to account for the range of column densities
observed in Fig \ref{FigPDR}.

\subsection{Comparison of observations with TDR model predictions}

\begin{figure*}[!ht]
\begin{center}
\includegraphics[width=13.0cm,angle=0]{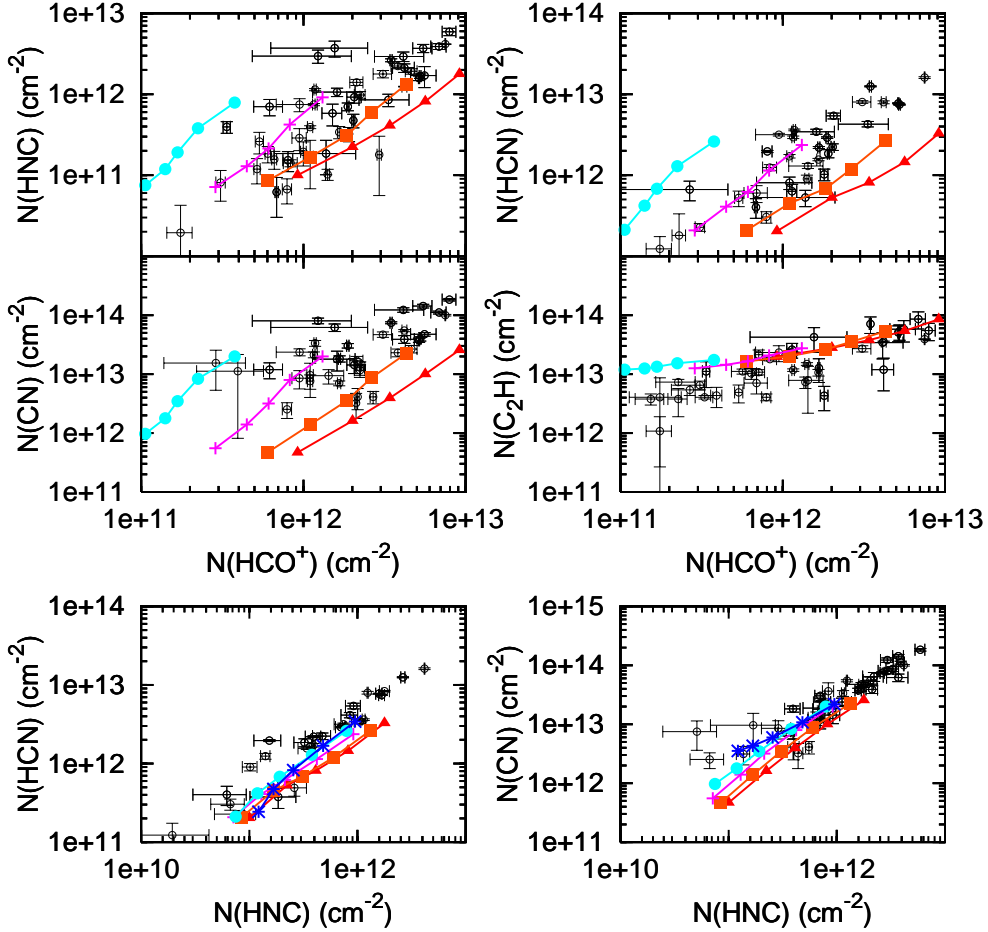}
\caption{Observations compared to the predictions of TDR models. The data
  (open circles) are from Lucas \& Liszt (2000), Liszt \& Lucas (2001), Gerin
  \etal (2010), and this work. The TDR models (filled symbols) are computed
  for several densities :  30 (triangles), 50 (squares), 100 (crosses), 200
  (circles), and 500 (double crosses) cm$^{-3}$. All models are computed for
  $N_{\rm H} = 1.8 \times 10^{21}$ cm$^{-2}$: note that if the total
  column density was larger, the model predictions would be simply multiplied
  by the same factor. The TDR models are computed for $A_V=0.5$ and $a$
  varying along each curve between $10^{-12}$ (top right) and $10^{-10}$
  s$^{-1}$ (bottom left).}
\label{FigRadio}
\end{center}
\end{figure*}

As suggested by Godard \etal (2009) and by the present observations, we
explore the TDR models in the range of parameters: $10^{-12} < a < 10^{-10}$
s$^{-1}$ and $30 < n_{\rm H} < 500$ cm$^{-2}$ for an ambient radiation field 
characterized by $\chi = 1$. Since the influence of this parameter on the
results of the PDR model (see Fig. \ref{FigPDR}) seems to be small
compared to the role of $N_{\rm H}$ and $n_{\rm H}$, we did not explore a
broader range of values for $\chi$.
The shielding from the ISRF is assumed uniform, 
$A_V = 0.5$ and the amount of gas sampled is fixed
to $N_{\rm H} = 1.8 \times 10^{21}$ cm$^{-2}$. Note that, in the TDR model,
$A_V$ is the actual UV-shielding of the gas and is no longer identical to the 
total hydrogen column density sampled by the line of sight (as it is the case
in the PDR model).

Fig. \ref{FigRadio} displays the comparison between the observed molecular
column densities and the predictions of the TDR models. Since the results
  of the TDR model are simply proportional to $N_{\rm H}$, the  computed
  values for higher (or lower) values of $N_{\rm H}$ are easily inferred
  from the display.
TDR models, without
fine tuning of the parameters, are consistent with the data. Over a broad
range of turbulent rates of strain $10^{-12}$ s$^{-1}$ $\leqslant a \leqslant$
$10^{-10}$ s$^{-1}$ and for a gas density 30 cm$^{-3}$ $\leqslant n_{\rm H}
\leqslant$ 200 cm$^{-3}$, the absolute column densities and column density
ratios of HCO$^{+}$, C$_2$H, CN, HCN, and HNC are reproduced.
The case $n_{\rm H} = 500$ cm$^{-3}$ is ruled out because the computed
HCO$^{+}$ column densities are too small compared to the observed values, even
if the total hydrogen column density $N_{\rm H}$ or the averaged turbulent
dissipative rate $\varepsilon$ are multiplied by a factor of 10.
For the CN-bearing
species, all deriving from HCNH$^+$ during the dissipation stage, the
agreement with the observations holds over a broader range of gas density : 30
cm$^{-3}$ $\leqslant n_{\rm H} \leqslant$ 500 cm$^{-3}$.
A trend, that was already found in Godard \etal (2009), is that the observed
column densities are in better agreement with models of small turbulent rate
of strain.



\section{Discussion}

\subsection{Grain surface chemistry} \label{GrainChemistry}

As shown in the previous section, if only the gas phase chemistry is
  taken into account, the
predictions of UV-dominated chemical models computed at moderate densities
($n_{\rm H} < 300$ cm$^{-3}$) are unable to explain the observed column
densities of 
the CN-bearing species. This result is in agreement with the previous
studies of Wagenblast \etal (1993), Crawford \& Williams (1997), and Liszt \&
Lucas (2001) which show that the predictions of purely gas phase PDR models
fail by a factor of 7 to account for the CN abundances observed in the
low density material toward $\zeta$ Per, and by factors of 30 and 40 to
reproduce the NH abundances observed toward $\zeta$ Per and $\zeta$ Oph
respectively.

In the present paper we explore the role of turbulent dissipation on the
cyanide chemistry that operates via the hydrogenation chain of ionized
carbon 
followed by the reaction chains (\ref{EqReacChain1}), (\ref{EqReacChain2}), and
(\ref{EqReacChain3}). An ongoing related work is the comparison of the
TDR model predictions to the Herschel-HIFI observations of nitrogen hydrides
along W31C and W49N (Persson \etal 2010).

But there is an alternative possibility to the pure gas phase chemistry.
It has been proposed that the grain surface chemistry dominate all the other
formation processes of the ammonia NH$_3$, and is therefore
responsible\footnote{ In a UV-dominated chemistry, the main destruction
    route of NH$_3$ is its photodissociation. The product NH$_2$ then leads to
    the formation of NH and CN (see Fig. \ref{FigChemistry0}).} for the
production of NH and CN in the diffuse ISM (van Dishoeck \& Black
1988). Assuming a NH$_3$ grain  formation rate coefficient of
\begin{equation}
k_G = 5.4 \times 10^{-18} \left( \frac{T\,\,\,[{\rm K}]}{m(N)\,\,\,[{\rm amu}]}
  \right)^{0.5} \quad {\rm cm}^{3} {\rm s}^{-1},
\end{equation}
corresponding to a surface reaction efficiency twice lower than that for the
H$_2$ formation on grains, Wagenblast \& Williams (1996) found that a UV
dominated chemical model reproduces the CN and NH abundances observed
toward $\zeta$ Per and $\zeta$ Oph. However, this approach is based on the
assumption that if a nitrogen atom strikes and sticks to a grain, ammonia is
quickly formed and comes off the grain. The resulting rate coeficient, namely
the cross section of the grain multiplied by the average speed of nitrogen
atoms, clearly sets an upper limit to the efficiency of the grain surface
chemistry.

\subsection{The UV radiation field}
Although, as said above, the comparison of model predictions to observations
is a difficult task because of the lack of information on the gas topology 
(density structure, division of the matter therefore shielding from the
ambient UV field, homogeneity of the gas conditions among the different
fragments) and the poorly known physical characteristics of what is
called a ``Gaussian-velocity component'',  we have three main constraints to
guide the models.

In the Gaussian-velocity components, {\it (i)} the gas density is lower than
$n_{\rm H}=5 \times 10^3$ \cc, {\it (ii)} where it is estimated, the column
density  of \HH\  is between 0.3 and 1.3 that of atomic hydrogen
(Table 4), and {\it (iii)} the {\it average} hydrogen  column density per unit
velocity interval ranges between  2 and 5 $10^{20}$ \cq/\kms. This implies
that the total hydrogen column density cannot be much larger than $4
  \times 10^{21}$ cm$^{-2}$ and that the  shielding from the ambient
radiation field is efficient (otherwise the \HH\ fraction  would be small).
According to PDR models, this imposes a narrow combination of the total column
density $N_{\rm H}$  and ambient UV field intensity $\chi$.
We are therefore confident that the gross properties of the UV illumination of
the gas in the  ``Gaussian velocity-components'' are well estimated.

\subsection{The velocity field} 

The fact that the widths of the \HCOp\ components are systematically larger
than those of the CN-bearing species suggests
an involvement of the velocity field in the
production and the evolution of these species. Observations in the visible
domain by Lambert \etal (1990), Crane \etal (1995), and Pan \etal (2004,
2005) already showed that the CH$^{+}$ line profiles are broader and less
Gaussian than those of CH, themselves broader than those of CN. It has been
proposed that 
such behaviours are due to spatial confinement of the molecules whose
production occurs hierarchically into denser and colder environments (Crawford
1995). But the differences observed among the linewidths 
may also bear the signatures of the dynamical processes at work in the
formation of all those molecules convolved with their most different
relaxation times (Godard \etal 2009).
Last, the similarity between the results obtained along Galactic lines of
sight and  in the Solar Neighbourhood also suggests a formation mechanism
weakly dependent  on the ambient UV field, i.e. possibly driven for
instance  by the dissipation of the ubiquitous turbulence. 

\section{Summary \& conclusions}

We have presented the analysis of single dish observations of mm absorption
lines of HCO$^{+}$, HNC, HCN, and CN toward remote star-forming regions in the
Galactic plane. The density of the gas responsible for the absorption lines 
is constrained by the excitation
temperature of the first rotational levels of HCO$^{+}$ and, in two cases, 
by line emission of CO (and isotopes). An upper limit $n_{\rm
  H} < 5 \times 10^{3}$ cm$^{-3}$ is obtained. The estimation of the total
column densities along the lines of sight, combined with their \HH\ fraction,
where known,  gives an upper limit on the UV-shielding per velocity
component,  $A_V < 2$. These conditions suggest that the  gas sampled along
the lines of sight belongs to the diffuse and/or translucent media. Because the
decomposition of the velocity structure into Gaussian components 
has been carried out recursively
for all the molecules and hyperfine components, 
intercomparisons of molecular abundances are feasible.
The inferred velocity structure will be helpful to unravel 
the analysis of future observations with Herschel/HIFI.

Our main result is the  similarity between the column densities per velocity
component derived in the Galactic plane gas (this work) and those derived in
the Solar Neighbourhood in previous works. 
It is unexpected because the spatial scales of the Galactic environment
sampled in this work are much larger. This suggests that the physical and
chemical processes involved in the formation and destruction of the analysed
species are similar in diffuse or translucent environments over the whole
Galactic disk.
We also find that the \HCOp\ linewidths are broader than those of
  the CN-bearing molecules.

We compare the observed column densities to two kinds of gas phase chemical
models, namely the photodissociation region (PDR)  and turbulent disspation
regions (TDR) models and find that: 
(1) the observed column densities of HCO$^{+}$ cannot be reproduced by PDR
models of diffuse or translucent gas, (2) the
column densities of CN-bearing species can be explained by PDR models
applied to dense ($n_{\rm H} > 300$ cm$^{-3}$) and well shielded ($A_V > 1$)
molecular gas, (3) the column densities and column density
ratios of HCO$^{+}$, CN, HCN, HNC, and C$_2$H are reproduced simultaneously by
TDR models of diffuse and translucent gas, over a broad range of turbulent
rates of strain ($10^{-12}$ s$^{-1}$ $\leqslant a \leqslant$ $10^{-10}$
s$^{-1}$), and gas densities (30 cm$^{-3}$ $\leqslant n_{\rm H} \leqslant$ 200
cm$^{-3}$). 

\acknowledgement
We are most grateful to Eric Herbst for the valuable information
  regarding the grain surface chemistry of nitrogen bearing species. We also
  thank the referee for providing constructive comments and help in improving
  the content of this paper.
\endacknowledgement

\appendix

\section{Gaussian decomposition and calculation of column densities} \label{AppendGauss}

Tabs. \ref{TabFitHCOp} - \ref{TabFitCN} contains the results of the Gaussian
decompostion procedure that we have applied to the spectra. The column
densities, given in the last columns, are derived assuming a single
excitation temperature $T_{\rm ex}$ for all the levels of a given molecule as
\begin{equation} \label{EqDcol}
N = Q(T_{\rm ex}) \frac{8 \pi \nu_0^{3}}{c^{3}} \frac{g_l}{g_u} \frac{1}{A_{ul}}
\left[1-e^{-h\nu_0/k T_{\rm ex}} \right]^{-1} \int \tau \, d\upsilon 
\end{equation}
where $\nu_0$, $g_u$, $g_l$ and $A_{ul}$ are the rest frequency, the upper and
lower level degeneracies and the Einstein's coefficients of the observed
transition, 
$Q(T_{\rm ex})$ is the partition funtion, and $c$ is the speed of light. We also remind
that for a Gaussian profile of peak opacity $\tau_0$ and FWHM $\Delta
\upsilon$, the opacity integral is $\int \tau \, d\upsilon = 1/2 \,
\sqrt{\pi/\ln{2}} \, \tau_0 \, \Delta \upsilon$. For an excitation temperature
of 2.73 K, Eq. \ref{EqDcol} becomes
\begin{equation}
N({\rm HCO}^{+}) = 1.10 \times 10^{12} \int \tau \, d\upsilon \quad {\rm cm}^{-2},
\end{equation}
\begin{equation}
N({\rm HNC})     = 1.78 \times 10^{12} \int \tau \, d\upsilon \quad {\rm cm}^{-2},
\end{equation}
\begin{equation}
N({\rm HCN})     = 3.43 \times 10^{12} \int \tau \, d\upsilon \quad {\rm cm}^{-2},
\end{equation}
and
\begin{equation}
N({\rm CN})      = 6.96 \times 10^{13} \int \tau \, d\upsilon \quad {\rm cm}^{-2}
\end{equation}
using the HCO$^{+}$ $J=0-1$, HNC $J=0-1$, HCN $J,F=0,1-1,2$, and CN
$J,F1,F=0,1/2,3/2-1,1/2,3/2$ transitions respectively.

\begin{table*}[]
\begin{center}
\caption{HCO$^{+}$ (0-1) absorption line analysis results.}
\begin{tabular}{l  c  c  c  c c}
\hline
 $^{a}$& $\upsilon_0$ & $\Delta \upsilon$ & $\tau_0$ & $N$(HCO$^{+}$) &$N_{\rm H}$ \\
Source      & (\kms) & (\kms) &     & ($10^{12}$ cm$^{-2}$)&($10^{20}$ cm$^{-2}$) \\
\hline
{\bf G05.88-0.39} &  -4.92  &1.13 $\pm$ 0.15& 0.09 $\pm$ 0.01 &  0.13 $\pm$  0.03 &  \\
\\
{\bf G08.67-0.36} &   9.82  &0.95 $\pm$ 0.12& 1.60 $\pm$ 0.28 &  2.02 $\pm$  0.58 &  \\
                  &  11.71  &1.40 $\pm$ 0.17& 1.65 $\pm$ 0.28 &  3.08 $\pm$  0.90 &  \\
                  &  14.52  &2.00 $\pm$ 0.18& 1.41 $\pm$ 0.18 &  3.74 $\pm$  0.83 &  \\
                  &  17.76  &1.17 $\pm$ 0.22& 0.87 $\pm$ 0.20 &  1.34 $\pm$  0.56 &  \\
                  &  20.00  &2.55 $\pm$ 0.63& 1.08 $\pm$ 0.16 &  3.65 $\pm$  1.45 &  \\
                  &  22.00  &1.02 $\pm$ 0.58& 0.24 $\pm$ 0.17 &  0.32 $\pm$  0.41 &  \\
\\
{\bf G10.62-0.38} &  11.62  &1.10 $\pm$ 0.12& 0.43 $\pm$ 0.04 &  0.62 $\pm$  0.12 &  \\
                  &  16.38  &2.35 $\pm$ 0.10& 2.23 $\pm$ 0.14 &  6.84 $\pm$  0.74 &  \\
                  &  19.00  &2.13 $\pm$ 0.20& 1.50 $\pm$ 0.09 &  4.18 $\pm$  0.64 &  \\
                  &  22.20  &2.70 $\pm$ 0.27& 1.57 $\pm$ 0.13 &  5.57 $\pm$  1.00 &  \\
                  &  22.95  &1.15 $\pm$ 0.45& 0.82 $\pm$ 0.18 &  1.23 $\pm$  0.75 &  \\
                  &  24.98  &1.65 $\pm$ 0.10& 0.70 $\pm$ 0.06 &  1.51 $\pm$  0.21 &  \\
                  &  27.50  &2.58 $\pm$ 0.17& 1.61 $\pm$ 0.11 &  5.46 $\pm$  0.74 &  \\
                  &  29.00  &1.53 $\pm$ 0.52& 0.78 $\pm$ 0.20 &  1.56 $\pm$  0.93 &  \\
                  &  30.50  &1.52 $\pm$ 0.32& 2.09 $\pm$ 0.27 &  4.13 $\pm$  1.39 &  \\
                  &  40.91  &2.60 $\pm$ 0.10& 2.33 $\pm$ 0.14 &  7.93 $\pm$  0.78 &  \\
                  &  44.94  &1.22 $\pm$ 0.05& 1.30 $\pm$ 0.07 &  2.06 $\pm$  0.21 &  \\
                  &  46.40  &0.60 $\pm$ 0.10& 0.26 $\pm$ 0.04 &  0.20 $\pm$  0.07 &  \\
\\
{\bf G34.3+0.1}   &   9.80  &0.53 $\pm$ 0.22& 0.38 $\pm$ 0.12 &  0.27 $\pm$  0.19 &  \\
                  &  10.60  &0.57 $\pm$ 0.12& 1.86 $\pm$ 0.61 &  1.38 $\pm$  0.72 &  \\
                  &  11.70  &1.37 $\pm$ 0.18& 1.82 $\pm$ 0.39 &  3.34 $\pm$  1.14 &  \\
                  &  14.07  &0.37 $\pm$ 0.08& 0.77 $\pm$ 0.23 &  0.38 $\pm$  0.20 &  \\
                  &  27.00  &1.02 $\pm$ 0.13& 1.19 $\pm$ 0.20 &  1.61 $\pm$  0.47 &  \\
                  &  27.75  &1.83 $\pm$ 0.32& 0.45 $\pm$ 0.09 &  1.10 $\pm$  0.41 &  \\
                  &  48.30  &2.23 $\pm$ 0.20& 0.94 $\pm$ 0.10 &  2.80 $\pm$  0.54 &  \\
\\
{\bf W49N}        &  33.20  &0.65 $\pm$ 0.03& 0.44 $\pm$ 0.02 &  0.34 $\pm$  0.02 & 3.0  \\
                  &  33.70  &2.23 $\pm$ 0.03& 0.63 $\pm$ 0.01 &  1.67 $\pm$  0.02 & 10.5 \\
                  &  36.50  &1.67 $\pm$ 0.12& 0.29 $\pm$ 0.01 &  0.60 $\pm$  0.03 & 7.8  \\
                  &  38.50  &1.75 $\pm$ 0.08& 0.91 $\pm$ 0.01 &  1.88 $\pm$  0.03 & 8.2  \\
                  &  39.70  &1.47 $\pm$ 0.02& 2.01 $\pm$ 0.03 &  3.47 $\pm$  0.05 & 6.9  \\
                  &  50.92  &1.65 $\pm$ 0.10& 0.35 $\pm$ 0.01 &  0.68 $\pm$  0.02 & 7.8  \\
                  &  53.40  &1.23 $\pm$ 0.10& 0.30 $\pm$ 0.02 &  0.43 $\pm$  0.03 & 5.8  \\
                  &  54.00  &3.38 $\pm$ 0.15& 0.51 $\pm$ 0.01 &  2.03 $\pm$  0.04 & 15.9 \\
                  &  56.70  &2.35 $\pm$ 0.07& 0.69 $\pm$ 0.01 &  1.91 $\pm$  0.03 & 11.0 \\
                  &  59.55  &2.35 $\pm$ 0.05& 1.89 $\pm$ 0.02 &  5.24 $\pm$  0.07 & 11.0 \\
                  &  60.80  &0.88 $\pm$ 0.05& 0.77 $\pm$ 0.05 &  0.80 $\pm$  0.06 & 4.1  \\
                  &  62.65  &2.95 $\pm$ 0.03& 1.48 $\pm$ 0.01 &  5.16 $\pm$  0.05 & 13.9 \\
                  &  66.21  &1.85 $\pm$ 0.07& 0.19 $\pm$ 0.01 &  0.41 $\pm$  0.01 & 8.7  \\
                  &  68.58  &1.80 $\pm$ 0.10& 0.24 $\pm$ 0.01 &  0.52 $\pm$  0.01 & 8.5  \\
                  &  71.11  &2.25 $\pm$ 0.20& 0.07 $\pm$ 0.00 &  0.19 $\pm$  0.01 & 10.6 \\
\\
{\bf W51}         &   3.80  &0.47 $\pm$ 0.12& 0.10 $\pm$ 0.02 &  0.06 $\pm$  0.03  & 1.0 \\
                  &   5.10  &1.23 $\pm$ 0.05& 0.72 $\pm$ 0.02 &  1.14 $\pm$  0.08  & 2.7 \\
                  &   6.25  &0.65 $\pm$ 0.03& 1.05 $\pm$ 0.06 &  0.87 $\pm$  0.10  & 1.4 \\
                  &   6.85  &2.01 $\pm$ 0.05& 0.70 $\pm$ 0.02 &  1.80 $\pm$  0.10  & 4.4 \\
                  &  12.71  &1.67 $\pm$ 0.18& 0.10 $\pm$ 0.01 &  0.20 $\pm$  0.04  & 3.7 \\
\hline
 $^{b}$     & $\upsilon_{\rm min}$ & $\upsilon_{\rm max}$ & $\int \tau d\upsilon$   &  $N$(HCO$^{+}$) & \\
Source      & (\kms) & (\kms) & (\kms) & ($10^{12}$ cm$^{-2}$) & \\
\hline
{\bf G10.62-0.38} & 13.6 & 15.2 &      0.77 $\pm$  0.08 &      0.94 $\pm$  0.10 & \\
                  & 31.0 & 33.0 &      3.77 $\pm$  0.31 &      4.63 $\pm$  0.38 & \\
                  & 33.3 & 34.7 &      1.81 $\pm$  0.15 &      2.23 $\pm$  0.18 & \\
                  & 34.7 & 36.0 &      1.74 $\pm$  0.16 &      2.13 $\pm$  0.19 & \\
                  & 36.0 & 37.5 &      3.10 $\pm$  0.29 &      3.82 $\pm$  0.36 & \\
                  & 37.5 & 40.0 & $>$  6.27         & $>$  7.71                & \\
\\
{\bf W49N}        & 41.0 & 47.0 &      0.90 $\pm$  0.03 &      1.00 $\pm$  0.04 & \\
\hline
\end{tabular}
\begin{list}{}{}
\item[$^{a}$] Multi-Gaussian decomposition. The column densities are derived
  assuming the excitation temperatures given in Table \ref{TabTex}.
\item[$^{b}$] Failure of the multi-gaussian decomposition. Column densities
  are inferred from the integral of the opacity assuming the excitation
  temperatures given in Table \ref{TabTex}.
\item[$^{c}$] Determined from the average total hydrogen column density per unit velocity
$N_{\rm H}/\Delta v = 2.2$ and 4.7$\times 10^{20}$ cm$^{-2}$/km s$^{-1}$
  toward W51 and W49N respectively.
\end{list}
\label{TabFitHCOp}
\end{center}
\end{table*}

\begin{table*}
\begin{center}
\caption{HNC (0-1) absorption line analysis products.}
\begin{tabular}{l  c  c  c   c }
\hline
 $^{a}$     & $\upsilon_0$         & $\Delta \upsilon$    & $\tau_0$& $N$(HNC)        \\
Source      & (\kms) & (\kms) &         & ($10^{12}$ cm$^{-2}$) \\
\hline
{\bf G10.62-0.38} &  11.62  &1.72 $\pm$ 0.22& 0.19 $\pm$ 0.02 &  0.72 $\pm$  0.17  \\
                  &  14.42  &1.23 $\pm$ 0.42& 0.09 $\pm$ 0.02 &  0.24 $\pm$  0.14  \\
                  &  16.38  &1.30 $\pm$ 0.05& 1.43 $\pm$ 0.07 &  4.02 $\pm$  0.33  \\
                  &  19.16  &2.01 $\pm$ 0.15& 0.50 $\pm$ 0.03 &  2.19 $\pm$  0.27  \\
                  &  22.20  &2.15 $\pm$ 0.28& 0.38 $\pm$ 0.06 &  1.76 $\pm$  0.51  \\
                  &  22.95  &1.17 $\pm$ 0.13& 1.21 $\pm$ 0.08 &  3.06 $\pm$  0.57  \\
                  &  24.98  &2.05 $\pm$ 0.35& 0.13 $\pm$ 0.02 &  0.60 $\pm$  0.18  \\
                  &  27.80  &1.50 $\pm$ 0.07& 1.18 $\pm$ 0.08 &  3.82 $\pm$  0.44  \\
                  &  29.10  &1.32 $\pm$ 0.18& 1.16 $\pm$ 0.10 &  3.84 $\pm$  0.85  \\
                  &  30.25  &1.13 $\pm$ 0.07& 1.23 $\pm$ 0.10 &  3.02 $\pm$  0.41  \\
                  &  31.90  &2.31 $\pm$ 0.37& 0.45 $\pm$ 0.03 &  2.28 $\pm$  0.50  \\
                  &  34.00  &1.08 $\pm$ 0.22& 0.32 $\pm$ 0.04 &  0.76 $\pm$  0.25  \\
                  &  35.35  &1.02 $\pm$ 0.13& 0.53 $\pm$ 0.05 &  1.16 $\pm$  0.26  \\
                  &  36.51  &1.22 $\pm$ 0.10& 0.81 $\pm$ 0.05 &  2.12 $\pm$  0.31  \\
                  &  38.80  &2.16 $\pm$ 0.18& 0.72 $\pm$ 0.03 &  3.38 $\pm$  0.45  \\
                  &  40.91  &1.30 $\pm$ 0.05& 2.16 $\pm$ 0.14 &  6.10 $\pm$  0.63  \\
                  &  44.94  &1.30 $\pm$ 0.10& 0.34 $\pm$ 0.03 &  0.95 $\pm$  0.15  \\
\\
{\bf G34.3+0.1}   &  10.60  &0.58 $\pm$ 0.13& 0.15 $\pm$ 0.03 &  0.19 $\pm$  0.08  \\
                  &  11.80  &1.45 $\pm$ 0.15& 0.28 $\pm$ 0.02 &  0.85 $\pm$  0.15  \\
                  &  27.00  &1.05 $\pm$ 0.07& 0.48 $\pm$ 0.03 &  1.06 $\pm$  0.13  \\
\\
{\bf W49N}        &  33.50  &1.77 $\pm$ 0.17& 0.12 $\pm$ 0.01 &  0.39 $\pm$  0.06  \\
                  &  38.50  &1.68 $\pm$ 0.10& 0.22 $\pm$ 0.01 &  0.70 $\pm$  0.08  \\
                  &  39.70  &0.95 $\pm$ 0.02& 1.46 $\pm$ 0.04 &  2.63 $\pm$  0.11  \\
                  &  50.92  &0.53 $\pm$ 0.15& 0.06 $\pm$ 0.01 &  0.06 $\pm$  0.03  \\
                  &  54.00  &2.85 $\pm$ 0.32& 0.09 $\pm$ 0.01 &  0.47 $\pm$  0.09  \\
                  &  56.70  &1.78 $\pm$ 0.18& 0.10 $\pm$ 0.01 &  0.32 $\pm$  0.06  \\
                  &  59.40  &2.01 $\pm$ 0.08& 0.44 $\pm$ 0.01 &  1.67 $\pm$  0.11  \\
                  &  60.90  &0.97 $\pm$ 0.13& 0.08 $\pm$ 0.01 &  0.15 $\pm$  0.04  \\
                  &  62.60  &2.15 $\pm$ 0.08& 0.39 $\pm$ 0.01 &  1.58 $\pm$  0.11  \\
\hline
 $^{b}$     & $\upsilon_{\rm min}$ & $\upsilon_{\rm max}$ & $\int \tau d\upsilon$   &  $N$(HNC) \\
Source      & (\kms) & (\kms) & (\kms) & ($10^{12}$ cm$^{-2}$) \\
\hline
{\bf G10.62-0.38} & 46.0 & 47.0 & $<$  0.03         & $<$  0.07                 \\
\\
{\bf G34.3+0.1}   &  9.0 & 10.1 & $<$  0.03         & $<$  0.06                 \\
                  & 13.3 & 14.7 & $<$  0.04         & $<$  0.07                 \\
                  & 46.0 & 50.0 & $<$  0.06         & $<$  0.11                 \\
\\
{\bf W49N}        & 35.0 & 37.0 &      0.09 $\pm$  0.02 &      0.16 $\pm$  0.04 \\
                  & 41.0 & 47.0 &      0.11 $\pm$  0.04 &      0.19 $\pm$  0.06 \\
                  & 65.0 & 67.3 &      0.10 $\pm$  0.02 &      0.18 $\pm$  0.04 \\
                  & 67.3 & 69.7 &      0.07 $\pm$  0.02 &      0.12 $\pm$  0.04 \\
                  & 69.7 & 72.5 & $<$  0.02         & $<$  0.04                 \\  
\\
{\bf W51}         &  3.0 &  5.5 &      0.09 $\pm$  0.05 &      0.17 $\pm$  0.10 \\
                  &  5.5 &  9.5 &      0.09 $\pm$  0.06 &      0.18 $\pm$  0.12 \\
                  & 11.0 & 14.0 & $<$  0.06         & $<$  0.11                 \\
\hline
\end{tabular}
\begin{list}{}{}
\item[$^{a}$] Same as Table \ref{TabFitHCOp}.
\item[$^{b}$] Same as Table \ref{TabFitHCOp}.
\end{list}
\label{TabFitHNC}
\end{center}
\end{table*}

\begin{table*}[]
\begin{center}
\caption{HCN (0-1) absorption line analysis products.}
\begin{tabular}{l  c  c  c  c }
\hline
 $^{a}$     & $\upsilon_0$         & $\Delta \upsilon$    & $\tau_0$ & $N$(HCN)        \\
Source      & (\kms) & (\kms) &          & ($10^{12}$ cm$^{-2}$) \\
\hline
{\bf G10.62-0.38} &  44.97  &0.92 $\pm$ 0.03& 1.45 $\pm$ 0.07 &  5.40 $\pm$  0.45  \\
\\
{\bf G34.3+0.1}   &  10.10  &1.00 $\pm$ 0.15& 0.16 $\pm$ 0.02 &  0.66 $\pm$  0.18  \\
                  &  10.60  &0.32 $\pm$ 0.03& 0.40 $\pm$ 0.04 &  0.53 $\pm$  0.12  \\
                  &  11.80  &1.22 $\pm$ 0.05& 0.85 $\pm$ 0.03 &  4.27 $\pm$  0.33  \\
                  &  27.00  &0.57 $\pm$ 0.02& 1.47 $\pm$ 0.06 &  3.44 $\pm$  0.24  \\
                  &  27.75  &0.55 $\pm$ 0.05& 0.35 $\pm$ 0.03 &  0.81 $\pm$  0.14  \\
\\
{\bf W49N}        &  33.40  &1.43 $\pm$ 0.05& 0.42 $\pm$ 0.01 &  2.20 $\pm$  0.06  \\
                  &  38.50  &1.22 $\pm$ 0.03& 0.65 $\pm$ 0.01 &  2.91 $\pm$  0.06  \\
                  &  39.70  &0.87 $\pm$ 0.02& 3.94 $\pm$ 0.06 & 12.51 $\pm$  0.18  \\
                  &  51.00  &1.00 $\pm$ 0.17& 0.11 $\pm$ 0.01 &  0.40 $\pm$  0.11  \\
                  &  54.00  &2.50 $\pm$ 0.20& 0.24 $\pm$ 0.01 &  2.23 $\pm$  0.21  \\
                  &  56.60  &1.83 $\pm$ 0.15& 0.28 $\pm$ 0.01 &  1.85 $\pm$  0.22  \\
                  &  59.40  &1.50 $\pm$ 0.05& 1.35 $\pm$ 0.02 &  7.44 $\pm$  0.10  \\
                  &  60.90  &0.83 $\pm$ 0.05& 0.64 $\pm$ 0.02 &  1.96 $\pm$  0.05  \\
                  &  62.55  &1.83 $\pm$ 0.05& 1.15 $\pm$ 0.01 &  7.73 $\pm$  0.09  \\
\\
{\bf W51}         &   5.00  &0.93 $\pm$ 0.12& 0.17 $\pm$ 0.02 &  0.63 $\pm$  0.14  \\
                  &   6.80  &1.27 $\pm$ 0.12& 0.20 $\pm$ 0.02 &  1.02 $\pm$  0.17  \\
\hline
 $^{b}$     & $\upsilon_{\rm min}$ & $\upsilon_{\rm max}$ & $\int \tau d\upsilon$   &  $N$(HCN) \\
Source      & (\kms) & (\kms) & (\kms) & ($10^{12}$ cm$^{-2}$) \\
\hline
{\bf W49N}        & 69.7 & 72.5 &      0.03 $\pm$  0.03 &      0.18 $\pm$  0.15 \\
\\
{\bf W51}         & 11.0 & 14.0 & $<$  0.04         & $<$  0.27                 \\
\hline
\end{tabular}
\begin{list}{}{}
\item[$^{a}$] Same as Table \ref{TabFitHCOp}.
\item[$^{b}$] Same as Table \ref{TabFitHCOp}.
\end{list}
\label{TabFitHCN}
\end{center}
\end{table*}

\begin{table*}[]
\begin{center}
\caption{CN (0-1) absorption line analysis products.}
\begin{tabular}{l  c  c  c  c }
\hline
 $^{a}$     & $\upsilon_0$         & $\Delta \upsilon$    & $\tau_0$ & $N$(CN)        \\
Source      & (\kms) & (\kms) &        & ($10^{13}$ cm$^{-2}$) \\
\hline
{\bf G10.62-0.38} &  11.62  &0.92 $\pm$ 0.18& 0.16 $\pm$ 0.02 &  1.19  $\pm$ 0.35 \\
                  &  16.30  &1.20 $\pm$ 0.03& 1.16 $\pm$ 0.03 &  11.10 $\pm$ 0.63 \\
                  &  19.16  &1.03 $\pm$ 0.10& 0.47 $\pm$ 0.02 &  3.91  $\pm$ 0.59 \\
                  &  22.20  &1.83 $\pm$ 0.00& 0.32 $\pm$ 0.02 &  4.75  $\pm$ 0.35 \\
                  &  22.90  &0.87 $\pm$ 0.05& 1.16 $\pm$ 0.05 &  8.04  $\pm$ 0.87 \\
                  &  27.80  &1.47 $\pm$ 0.05& 1.23 $\pm$ 0.04 & 14.34  $\pm$ 1.05 \\
                  &  29.10  &0.78 $\pm$ 0.07& 0.99 $\pm$ 0.07 &  6.20  $\pm$ 0.92 \\
                  &  30.20  &1.10 $\pm$ 0.05& 1.39 $\pm$ 0.06 &  12.28 $\pm$ 1.05 \\
                  &  31.80  &2.10 $\pm$ 0.28& 0.38 $\pm$ 0.02 &  6.43  $\pm$ 1.12 \\
                  &  34.00  &1.22 $\pm$ 0.28& 0.19 $\pm$ 0.02 &  1.88  $\pm$ 0.63 \\
                  &  35.35  &0.68 $\pm$ 0.07& 0.44 $\pm$ 0.03 &  2.38  $\pm$ 0.38 \\
                  &  36.61  &1.40 $\pm$ 0.13& 0.50 $\pm$ 0.02 &  5.56  $\pm$ 0.77 \\
                  &  38.80  &1.63 $\pm$ 0.10& 0.66 $\pm$ 0.02 &  8.65  $\pm$ 0.84 \\
                  &  40.75  &1.35 $\pm$ 0.03& 1.70 $\pm$ 0.05 & 18.30  $\pm$ 0.95 \\
                  &  44.94  &0.83 $\pm$ 0.08& 0.22 $\pm$ 0.04 &  1.43  $\pm$ 0.41 \\
\\
{\bf G34.3+0.1}   &  27.00  &0.68 $\pm$ 0.12& 0.86 $\pm$ 0.17 &  1.79  $\pm$ 0.66 \\
\\
{\bf W49N}        &  33.30  &1.83 $\pm$ 0.12& 0.13 $\pm$ 0.01 &  1.82  $\pm$ 0.22 \\
                  &  38.40  &1.77 $\pm$ 0.08& 0.23 $\pm$ 0.01 &  3.02  $\pm$ 0.27 \\
                  &  39.60  &0.85 $\pm$ 0.02& 1.18 $\pm$ 0.03 &  7.43  $\pm$ 0.29 \\
                  &  59.20  &1.98 $\pm$ 0.08& 0.27 $\pm$ 0.01 &  3.98  $\pm$ 0.34 \\
                  &  62.40  &1.62 $\pm$ 0.07& 0.32 $\pm$ 0.01 &  3.77  $\pm$ 0.31 \\
\hline
 $^{b}$     & $\upsilon_{\rm min}$ & $\upsilon_{\rm max}$ & $\int \tau d\upsilon$   &  $N$(CN) \\
Source      & (\kms) & (\kms) & (\kms) & ($10^{13}$ cm$^{-2}$) \\
\hline
{\bf G10.62-0.38} & 13.6 & 15.2 &      0.18 $\pm$  0.06 &      0.85  $\pm$ 0.29 \\
                  & 24.0 & 26.0 &      0.26 $\pm$  0.07 &      1.29  $\pm$ 0.34 \\
                  & 46.0 & 47.0 & $<$  0.04         & $<$  0.21 \\
\\
{\bf G34.3+0.1}   &  9.0 & 10.1 &      0.20 $\pm$  0.13 &      1.54  $\pm$ 1.01 \\
                  & 10.1 & 11.0 & $<$  0.11         & $<$  0.84 \\
                  & 11.0 & 13.0 &      0.48 $\pm$  0.18 &      3.63  $\pm$ 0.14 \\
                  & 13.3 & 14.7 &      0.15 $\pm$  0.14 &      1.12  $\pm$ 1.05 \\
                  & 46.0 & 50.0 & $<$  0.19         & $<$ 1.46  \\
\\
{\bf W49N}        & 35.0 & 37.0 & $<$  0.03         & $<$  0.13 \\
                  & 41.0 & 47.0 & $<$  0.05         & $<$  0.22 \\
                  & 50.0 & 52.0 & $<$  0.03         & $<$  0.13 \\
                  & 52.5 & 55.5 &      0.09 $\pm$  0.04 &      0.41  $\pm$ 0.17 \\
                  & 55.5 & 57.9 &      0.07 $\pm$  0.03 &      0.32  $\pm$ 0.14 \\
                  & 60.3 & 61.3 &      0.09 $\pm$  0.02 &      0.41  $\pm$ 0.10 \\
\\
{\bf W51}         &  3.0 &  5.5 &      0.13 $\pm$  0.08 &      0.97  $\pm$ 0.56 \\
                  &  5.5 &  9.5 & $<$  0.09         & $<$  0.69 \\
                  & 11.0 & 14.0 & $<$  0.08         & $<$  0.59 \\
\hline
\end{tabular}
\begin{list}{}{}
\item[$^{a}$] Same as Table \ref{TabFitHCOp}.
\item[$^{b}$] Same as Table \ref{TabFitHCOp}.
\end{list}
\label{TabFitCN}
\end{center}
\end{table*}

\section{Impact of the abscissa uncertainty on the multi-Gaussian
  decomposition procedure} \label{AppendAbscissa}

In a spectrum, the possible velocity substructures are systematically
erased due to the finite velocity resolution $\delta \upsilon$. This could be
modelled as an uncertainty on the velocity position of each point.
Unfortunately the errors on the abscissa are rarely included in non-linear fitting
procedures because the system is considerably heavier to solve and because it
often prevents convergence. To evaluate the resulting uncertainties on the fit
parameters, namely the central opacity, the velocity centroid, and the FWHM,
we apply the fitting procedure on 3000 synthetic spectra of FWHM $\Delta
\upsilon$ varying between between 0.3 and 3.4 \kms\ sampled with the finite
spectral resolution of the observations $\delta \upsilon \sim 0.13$ \kms. A 
noise is added to the $x$-coordinates of all the spectral points. The rms of all
the measured linewidths $\Delta \upsilon'$ is found to scale as
\begin{equation}
\frac{\delta (\Delta \upsilon')}{\Delta \upsilon} = 0.45 \left( 
\frac{\delta \upsilon}{\Delta \upsilon} \right)^{3/2} 
\end{equation}
and decreases from 0.04 to 0.01 \kms\ as the true linewidth increases from
0.3 to 3.4 \kms. These uncertainties are smaller than (or comparable to) those
inferred from the fitting procedure and the resulting errors on the column
densities are at most 12\%. In comparison the resulting errors on central
opacities and velocity centroids are negligible.

\section{Cyanides chemical network} \label{AppendChemistry}

\begin{table*}[!h]
\begin{center}
\caption{Rates $k$ of the main reactions of the cyanide chemistry. The
photoreaction rates writte $k = \alpha \cdot {\rm exp}(-\gamma A_V)$ while all
the other reaction rates are derived as $k = \alpha \cdot (T/{\rm 300
K})^{\beta} {\rm exp}(-\gamma / T)$ where $T$ is the effective 
temperature (Zsarg\'o \& Federman 2003). Numbers in parenthesis are power of
10.} 
\begin{tabular}{l @{\hspace{0.1cm}} l @{\hspace{0.1cm}} l @{\hspace{0.1cm}} l
    c  @{\hspace{0.2cm}} c  @{\hspace{0.2cm}} c c}
\hline
\multicolumn{4}{c}{Reaction}   & $\alpha$ & $\beta$ & $\gamma$ & ref \\                    
\hline
CN            & + h$\nu$  & $\rightarrow$ C             & + N     & 4.64 (-10) &  0.00 & 3.6 & a \\
HCN           & + h$\nu$  & $\rightarrow$ CN            & + H     & 5.70 (-10) &  0.00 & 2.6 & a \\
HNC           & + h$\nu$  & $\rightarrow$ CN            & + H     & 5.70 (-10) &  0.00 & 2.6 & a \\
HCNH$^{+}$    & + e$^{-}$ & $\rightarrow$ HCN           & + H     & 9.45 (-07) & -0.65 & 0.0 & b,c \\
HCNH$^{+}$    & + e$^{-}$ & $\rightarrow$ HNC           & + H     & 9.45 (-07) & -0.65 & 0.0 & b,c \\
HCNH$^{+}$    & + e$^{-}$ & $\rightarrow$ CN            & + 2H    & 9.10 (-08) & -0.65 & 0.0 & b,c \\
HCNH$^{+}$    & + e$^{-}$ & $\rightarrow$ CN            & + H$_2$ & 1.80 (-08) & -0.65 & 0.0 & b,c \\
H$_2$NC$^{+}$ & + e$^{-}$ & $\rightarrow$ HNC           & + H     & 1.75 (-07) & -0.50 & 0.0 & d \\
H$_2$NC$^{+}$ & + e$^{-}$ & $\rightarrow$ CN            & + H$_2$ & 1.75 (-07) & -0.50 & 0.0 & d \\
HNC$^{+}$     & + e$^{-}$ & $\rightarrow$ CN            & + H     & 3.90 (-07) & -1.00 & 0.0 & e,f \\
HCN$^{+}$     & + e$^{-}$ & $\rightarrow$ CN            & + H     & 3.90 (-07) & -1.00 & 0.0 & e \\
CH$_3$$^{+}$  & + N       & $\rightarrow$ HCN$^{+}$     & + H$_2$ & 2.20 (-11) &  0.00 & 0.0 & g \\
CH$_3$$^{+}$  & + N       & $\rightarrow$ HCNH$^{+}$    & + H     & 2.20 (-11) &  0.00 & 0.0 & g \\
CH$_3$$^{+}$  & + N       & $\rightarrow$ H$_2$NC$^{+}$ & + H     & 2.20 (-11) &  0.00 & 0.0 & g \\
CH$_2$$^{+}$  & + N       & $\rightarrow$ HCN$^{+}$     & + H     & 4.70 (-10) &  0.00 & 0.0 &  \\
CH$_2$$^{+}$  & + N       & $\rightarrow$ HNC$^{+}$     & + H     & 4.70 (-10) &  0.00 & 0.0 &  \\
CH$^{+}$      & + N       & $\rightarrow$ CN$^{+}$      & + H     & 1.90 (-10) &  0.00 & 0.0 & h \\
CH$_2$        & + N       & $\rightarrow$ HCN           & + H     & 3.95 (-11) &  0.17 & 0.0 & i \\
CH            & + N       & $\rightarrow$ CN            & + H     & 2.00 (-10) &  0.00 & 0.0 & j \\
C$^{+}$       & + NH      & $\rightarrow$ CN$^{+}$      & + H     & 7.80 (-10) &  0.00 & 0.0 &  \\
C$^{+}$       & + NH$_2$  & $\rightarrow$ HCN$^{+}$     & + H     & 1.10 (-09) &  0.00 & 0.0 &  \\
HCN$^{+}$     & + H$_2$   & $\rightarrow$ HCNH$^{+}$    & + H     & 9.00 (-10) &  0.00 & 0.0 & k \\
HNC$^{+}$     & + H$_2$   & $\rightarrow$ HCNH$^{+}$    & + H     & 9.00 (-10) &  0.00 & 0.0 & k \\
\hline
\end{tabular}
\begin{list}{}{}
\item[$^{a}$] Roberge \etal (1991).
\item[$^{b}$] Florescu-Mitchell \& Mitchell (2006).
\item[$^{c}$] Semaniak \etal (2001).
\item[$^{d}$] Adams \& Smith (1988).
\item[$^{e}$] Sheehan \etal (1999).
\item[$^{f}$] Talbi \etal (2000).
\item[$^{g}$] Fehsenfeld (1976).
\item[$^{h}$] Viggiano \etal (1980).
\item[$^{i}$] Smith \etal (2004).
\item[$^{j}$] Brownsword \etal (1996).
\item[$^{k}$] Huntress (1977).
\end{list}
\label{TabChem}
\end{center}
\end{table*}

\begin{figure*}[]
\begin{center}
\includegraphics[width=14.cm,angle=0]{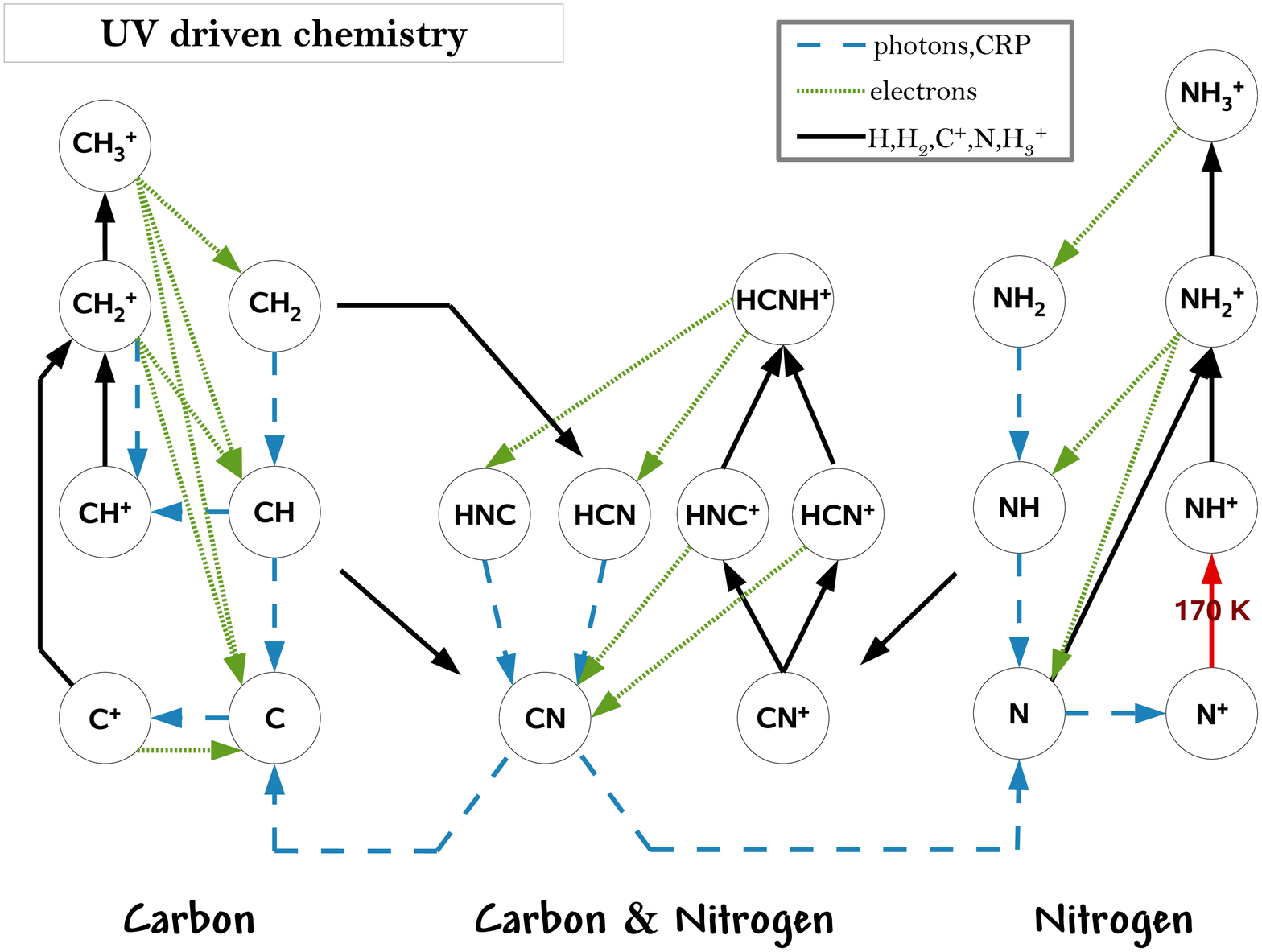}
\caption{Chemical network of a UV-dominated chemistry: $n_{\rm H}$ = 50
  cm$^{-3}$ and $A_V$ = 0.4. This figure is simplified: for each species, only the
  reactions which altogether contribute at least to 70 percent of the total
  destruction and formation rate are displayed. The red arrow show the
  endoenergetic reactions with the energy involved.}
\label{FigChemistry0}
\vspace{0.5cm}
\includegraphics[width=14.cm,angle=0]{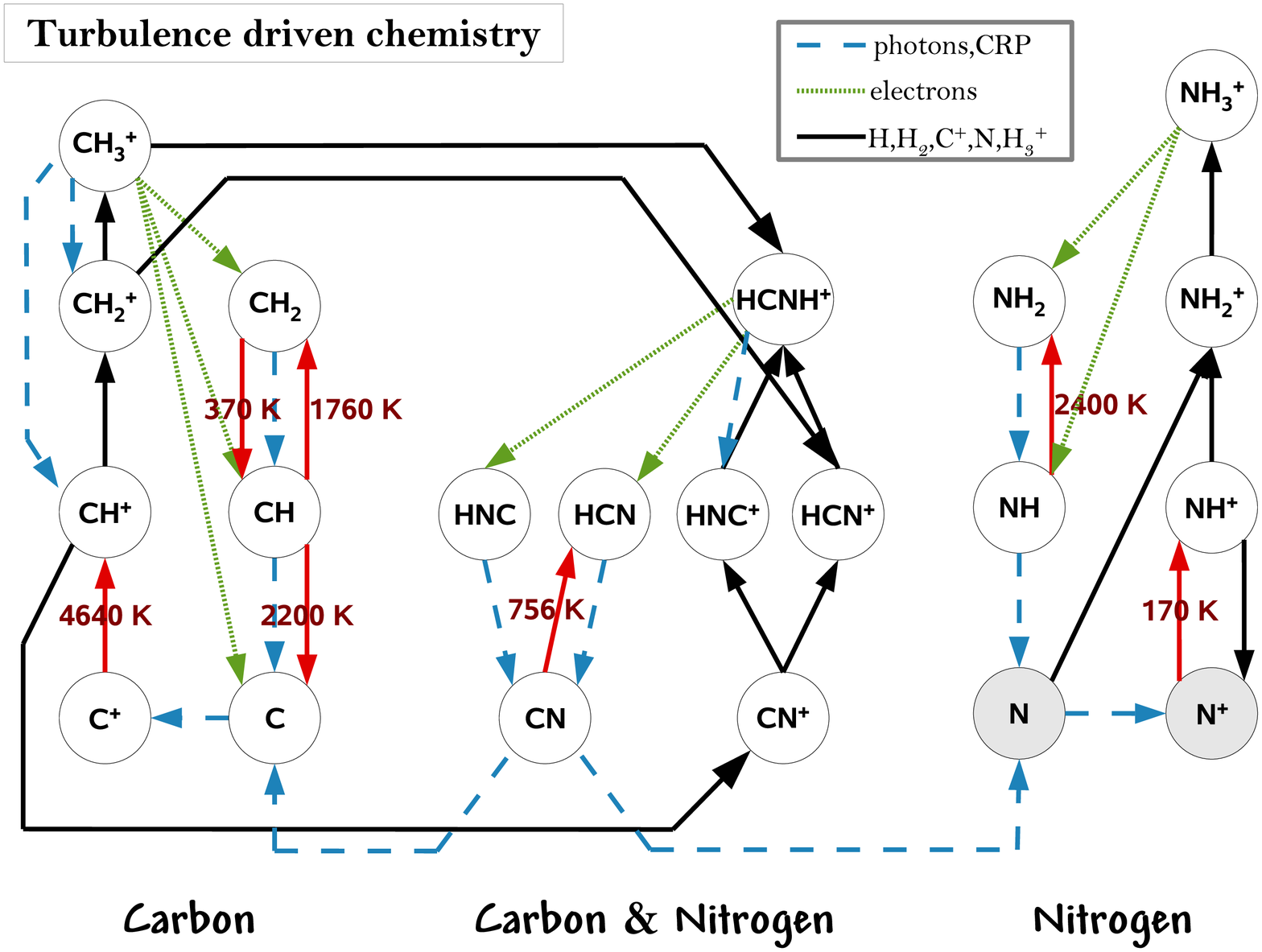}
\caption{Same as Fig. \ref{FigChemistry0} for a turbulence-dominated chemistry:
  $n_{\rm H}$ = 50 cm$^{-3}$, $A_V$ = 0.4 and $a = 10^{-11}$ s$^{-1}$. For the
  sake of simplicity, the main destruction route of N (N + CH$_2^{+}$
  $\rightarrow$ HCN$^{+}$ + H) and the main formation pathway of N$^{+}$
  (photodissociation of NO$^{+}$) are not displayed here: those two species
  are therefore highlighted.}
\label{FigChemistry}
\end{center}
\end{figure*}


Figs. \ref{FigChemistry0} \&  \ref{FigChemistry} show the main production
  and destruction pathways of the hydrogenation chains of carbon, nitrogen, and
cyano, resulting from the PDR ($n_{\rm H} = 50$ cm$^{-3}$, $A_V = 0.4$) and
TDR ($n_{\rm H} = 50$ cm$^{-3}$, $A_V = 0.4$, $a = 10^{-11}$ s$^{-1}$) models
respectively. These figures are simplified: for each species, only the
  reactions which altogether contribute at least to 70 percent of the total
  destruction and formation rate are displayed. There is one major difference
  between these networks: in a UV-dominated chemical model, the cyanide
  chemistry is initiated by: 
  \begin{equation}
    {\rm CH}   + {\rm N} \rightarrow {\rm CN} + {\rm H},
  \end{equation}
  \begin{equation}
    {\rm CH}_2 + {\rm N} \rightarrow {\rm HCN} + {\rm H},
  \end{equation}
  and
  \begin{equation}
    {\rm NH} + {\rm C}^{+} \rightarrow {\rm CN}^{+} + {\rm H} ;
  \end{equation}
  while in a chemistry driven by turbulent dissipation, the hydrogenation
  chain of cyano is triggered by the ion-neutral reactions: 
  \begin{equation}
    {\rm CH}^{+}   + {\rm N} \rightarrow {\rm CN}^{+}   + {\rm H},
  \end{equation}
  \begin{equation}
    {\rm CH}_2^{+} + {\rm N} \rightarrow {\rm HCN}^{+}  + {\rm H},
  \end{equation}
  and
  \begin{equation}
    {\rm CH}_3^{+} + {\rm N} \rightarrow {\rm HCNH}^{+} + {\rm H}.
  \end{equation}
 Since the pathways displayed in Figs. \ref{FigChemistry0} \&
  \ref{FigChemistry} depend on
  the chemical rates, and since the nitrogen and cyanide chemistry are still
  poorly known, we list the chemical rates we have adopted in our models for
  several reactions in Table \ref{TabChem}.

\end{document}